\documentclass[superscriptaddress,
reprint,nofootinbib,
preprintnumbers,
amsmath,amssymb,aps]{revtex4-1}

\pdfoutput=1
\usepackage{balance}
\usepackage[utf8]{inputenc} % To support utf8 characters
\usepackage[T1]{fontenc} % T1 font encoding
\usepackage{microtype}
\usepackage{verbatim}
\usepackage{amsmath,comment}
\usepackage{mathtools}

\usepackage{graphicx}% Include figure files

\usepackage[dvipsnames]{xcolor}

\usepackage{dcolumn}% Align table columns on decimal point
\usepackage{bm}% bold math
\usepackage[colorlinks = true,
            linkcolor = blue,
            urlcolor  = blue,
            citecolor =red,
            anchorcolor = blue]{hyperref}
\usepackage{verbatim}
\usepackage{color,ulem}
\usepackage[english]{babel}

\usepackage{blindtext}
\usepackage{float,amsmath,amssymb}

\newcommand{\dfr}[2]{\frac {\displaystyle #1}{\displaystyle #2}}

\usepackage{graphicx}
\usepackage{amsmath}
\usepackage{gensymb}
\usepackage{physics}
\usepackage{balance}
\usepackage[varg]{txfonts}
\usepackage{ulem,lipsum}
\usepackage{fancyhdr}

\begin{document}

\preprint{IFT-UAM/CSIC-20-44}
\title{Field Theory of Dissipative Systems with Gapped Momentum States}
\author{M. Baggioli}
\email{matteo.baggioli@uam.es}
\affiliation{Instituto de Fisica Teorica UAM/CSIC, c/ Nicolas Cabrera 13-15, Cantoblanco, 28049 Madrid, Spain}
\author{M. Vasin}
%\email{professorvasin@gmail.com}
\affiliation{Institute for High Pressure Physics, RAS, 142190, Moscow, Russia}
\author{V. V. Brazhkin}
%\email{brazhkin@hppi.troitsk.ru}
\affiliation{Institute for High Pressure Physics, RAS, 142190, Moscow, Russia}
\author{K. Trachenko}
\email{k.trachenko@qmul.ac.uk}
\affiliation{School of Physics and Astronomy, Queen Mary University of London, Mile End Road, London, E1 4NS, UK}

\begin{abstract}
We develop a field theory with dissipation based on a finite range of wave propagation and associated gapped momentum states in the wave spectrum. We analyze the properties of the Lagrangian and the Hamiltonian with two scalar fields in different representations and show how the new properties of the two-field Lagrangian are related to Keldysh-Schwinger formalism. The proposed theory is non-Hermitian, and we discuss its properties related to $\mathcal{PT}$ symmetry. The calculated correlation functions show a decaying oscillatory behavior related to gapped momentum states. We corroborate this result using path integration. The interaction potential becomes short-ranged due to dissipation. Finally, we observe that the proposed field theory represents a departure from the harmonic paradigm and discuss the implications of our theory for the Lagrangian formulation of hydrodynamics. 
\end{abstract}

{\let\clearpage\relax
\maketitle
}

\tableofcontents

\section{Introduction}

The basic assumptions and results of statistical physics are related to introducing, and frequently exploiting, the concept of a closed or quasi-closed system or subsystem. This includes, for example, most of the central ideas of a statistical or thermal equilibrium ensuing the definitions of entropy and  temperature, the statistical independence of the subsystems and the consequential additivity of the logarithm of the statistical distribution function \cite{landau}. 

A closed system is an approximation simplifying a theoretical description. This approximation does not apply in several important cases, including in small systems that have been of interest in the area of condensed matter recently or in systems with short relaxation time. In this case, a theory needs to deal with an {\it open} system. 

On general grounds, the theoretical description of open systems and dissipation is an interesting and challenging problem related to finding new concepts and ideas. In quantum-mechanical systems, this problem is viewed as a core problem in modern physics \cite{rotter2015review} and is related to the foundations of quantum theory itself (see, e.g., Refs. \cite{rotter2015review,benderbook,mohsen,bender,rotter1}). Describing dissipation has seen renewed recent interest in areas related to non-equilibrium and irreversible physics, decoherence effects, complex systems and hydrodynamics \cite{Glorioso:2018wxw}. 

A related conceptually difficult problem is to describe an open system using a field theory based on a Lagrangian and to account for dissipation and irreversibility (see, e.g., Refs. \cite{kamenev,benderbook,endlich,crossley} and references therein). 

Starting from early work (see, e.g., \cite{feynman,leggett}), a common approach to treat dissipation is to introduce a central dissipative system of interest together with its environment modelled as, for example, a bath of harmonic oscillators and an interaction between the two sectors enabling energy exchange (see, e.g., Refs. \cite{kamenev,weiss} for review). In this picture, dissipative effects can be discussed by solving simplified models exploiting approximations such as the linearity of the system and its couplings. 

Another approach relies on holographic techniques \cite{Baggioli:2019rrs}, where dissipation is encoded into the black hole horizon dynamics. These methods have been very fruitful in describing strongly coupled dissipative fluids \cite{Policastro:2002se,Janik:2006ft,Baggioli:2018vfc,Baggioli:2018nnp,Baggioli:2019aqf,Baggioli:2019sio}. However, the nature of the dual field theory is not easily accessible and is often obscure.

Here, we develop a field theory which describes dissipation based on a conceptually different idea. We do not consider an explicit coupling between a central system with its environment. Instead, the idea is based on the dissipation of an excitation which is {\it not} an eigenstate in the system where it propagates. This approach draws on recent understanding of phonon propagation in liquids and associated dissipation of these phonons \cite{ropp,prl,pre}. 

No dissipation takes place when a plane wave propagates in an ideal crystal where the wave is an eigenstate. However, a plane wave dissipates in systems with structural and dynamical disorder such as glasses, liquids or other systems with strong anharmonicities because its not an eigenstate in those systems \cite{collective} (see, e.g., Refs. \cite{PhysRevLett.122.145501,PhysRevE.100.062131,PhysRevResearch.1.012010,PhysRevResearch.2.013267} for recent field theory applications of related effects in amorphous and disordered systems). 

We note that the overall system (liquid in our case) is conservative and does not lose or gain energy. Similarly, collective excitations in a disordered system such as glass or liquid, generally defined as eigenstates in that system, do not decay either \cite{collective}. (We note that calculating these excitations presents an exponentially complex problem because it involves a large number of strongly-coupled non-linear oscillators \cite{ropp}.) In our consideration, a decaying object that loses energy is the harmonic plane wave (phonon) propagating in a disordered system where its {\it not} an eigenstate. Indeed, plane waves constantly evolve in a disordered system, e.g. decay into other waves and emerge anew due to thermal fluctuations. Accordingly, an open system in our consideration is the harmonic collective excitation, the plane wave, operating in a system where its not an eigenstate.

An important effect related to wave dissipation is the emergence of the gap in $k$- or momentum space in the transverse wave spectrum, with the accompanying decrease of the wave energy due to dissipation \cite{ropp,prl,pre}. This gap (a) shrinks the range of $k$-vectors where phonons can propagate and (b) reduces the energy of the remaining propagating phonons. 

It has been realized that in addition to liquids, gapped momentum states (GMS) emerge in a surprising variety of areas \cite{Baggioli:2019jcm}, including strongly-coupled plasma, electromagnetic waves, non-linear Sine-Gordon model, relativistic hydrodynamics and holographic models.

The field theory developed here describes dissipation on the basis of GMS. GMS is a well-specified effect and is naturally suited for describing the dissipation in a field theory because the field theory is deeply rooted in the harmonic paradigm involving the propagation of plane waves \cite{zee}. Despite its specificity, this effect and the proposed field theory are generally applicable to a wide range of physical phenomena in interacting systems where collective excitations propagate. \\

In the next section \ref{frenk}, we briefly review the emergence of gapped momentum states in Maxwell-Frenkel theory and their properties. We subsequently recall the two-field Lagrangian which gives rise to gapped momentum states and expand on different formulations, solutions and properties of this Lagrangian in section \ref{sectwofields}. This includes the discussion of the Hamiltonian and energy and their lower bounds in different regimes. We also show how two new properties of the two-field Lagrangian are related to results from Keldysh-Schwinger formalism in section \ref{six}. In the following section \ref{secherm}, we address the non-Hermiticity of the proposed Lagrangian and Hamiltonian operators as well as their $\mathcal{PT}$ symmetry and its breaking. We calculate the correlation functions in section \ref{seccorr} and find that they show a decayed oscillatory behavior, with frequency and decay related to GMS. We corroborate this result using path integration. In section \ref{secint}, we observe that the interaction potential becomes short-ranged due to dissipation. Finally, we discuss other implications of the proposed field theory, including the ways in which it departs from the harmonic paradigm (section \ref{secharm}) and its implications for the field-theoretical description of hydrodynamics (section \ref{sechydro}). 

\section{Dissipation and gapped momentum states in the Maxwell-Frenkel theory}\label{frenk}

We start with recalling how liquid transverse modes develop gapped momentum states (GMS) and how this effect can be represented by a Lagrangian. We note that a first-principles description of liquids is exponentially complex and is not tractable because it involves a large number of coupled non-linear oscillators \cite{ropp}. At the same time, liquids have no simplifying small parameters as in gases and solids \cite{landau}. However, progress in understanding liquid modes can be made by using a non-perturbative approach to liquids pioneered by Maxwell and developed later by Frenkel. This program involves the Maxwell interpolation:
\begin{equation}
\frac{ds}{dt}=\frac{P}{\eta}+\frac{1}{G}\frac{dP}{dt}
\label{a1}
\end{equation}
\noindent where $s$ is shear strain, $\eta$ is viscosity, $G$ is shear modulus and $P$ is shear stress.

Eq. (\ref{a1}) reflects Maxwell's proposal \cite{maxwell} that shear response in a liquid is the sum of viscous and elastic responses given by the first and second right-hand side . Eq. (\ref{a1}) serves as the basis of liquid viscoelasticity.

Frenkel proposed \cite{frenkel} to represent the Maxwell interpolation (\ref{a1}) by introducing the operator $A=1+\tau\frac{d}{dt}$ and write Eq. (\ref{a1}) as $\frac{ds}{dt}=\frac{1}{\eta}AP$. Here, $\tau$ is the Maxwell relaxation time $\frac{\eta}{G}$. Frenkel's theory has identified $\tau$ with the average time between consecutive molecular jumps in the liquid \cite{frenkel}. This has become an accepted view \cite{dyre}. Frenkel's next idea was to generalize $\eta$ to account for liquid's short-time elasticity as $\frac{1}{\eta}\rightarrow\frac{1}{\eta}\left(1+\tau\frac{d}{dt}\right)$ and use this $\eta$ in the Navier-Stokes equation $\nabla^2{\bf v}=\frac{1}{\eta}\rho\frac{d{\bf v}}{dt}$, where ${\bf v}$ is velocity, $\rho$ is density and $\frac{d}{dt}=\frac{\partial}{\partial t}+{\bf v\nabla}$. We have carried this idea forward \cite{ropp} and, considering small ${\bf v}$, wrote:
\begin{equation}
c^2\frac{\partial^2v}{\partial x^2}=\frac{\partial^2v}{\partial t^2}+\frac{1}{\tau}\frac{\partial v}{\partial t}
\label{gener3}
\end{equation}

\noindent where $v$ is the velocity component perpendicular to $x$, $\eta=G\tau=\rho c^2\tau$ and $c$ is the shear wave velocity.

In contrast to the Navier-Stokes equations, Eq.(\ref{gener3}) contains the second time derivative and hence gives propagating waves. We solved Eq. (\ref{gener3}) in Ref. \cite{ropp}: seeking the solution as $v=v_0\exp\left(i(kx-\omega t)\right)$ gives \begin{equation}
    \omega^2+\omega\frac{i}{\tau}-c^2k^2=0
\end{equation}
with the complex solutions
\begin{equation}
    \omega=-\frac{i}{2\tau}\pm\sqrt{c^2k^2-\frac{1}{4\tau^2}}\label{compl}
\end{equation}

\noindent where we assumed $\omega$ to be complex and $k$ real, corresponding to time decay.

Then, $v$ can be written as
\begin{equation}
v\propto\,e^{-\frac{t}{2\tau}}\,e^{i\,\omega_Rt}
\label{distau}
\end{equation}

\noindent where
\begin{equation}
\omega_R=\sqrt{c^2k^2-\frac{1}{4\tau^2}}
\label{omega}
\end{equation}

The absence of dissipation in (\ref{gener3}) corresponds to setting $\tau\rightarrow\infty$ in (\ref{omega}) and to an infinite range of propagation of the plane shear wave, as in the a ideal ordered crystal. A finite $\tau$ implies dissipation of the wave in a sense that it acquires a {\it finite} propagation range. Indeed, the dissipation takes place over time approximately equal to $\tau$ according to (\ref{distau}). $\tau$ sets the physical time scale during which we consider the dissipation process: if an observation of an injected shear wave starts at $t=0$, time $t\approx\tau$ is the end of the process because over this time the wave amplitude and energy appreciably reduce.

An important property is the emergence of the gap in $k$-space or GMS: in order for $\omega$ in (\ref{omega}) to be real, $k>k_g$ should hold, where
\begin{equation}
k_g=\frac{1}{2c\tau}
\label{kgap}
\end{equation}

\noindent increases with temperature because $\tau$ decreases.

The value of $k$-gap (\ref{kgap}) is related to the finite propagation range of transverse waves. Indeed, if $\tau$ is the time during which a shear stress can exist in a liquid, the liquid elasticity length $d_{\rm el}=c\tau$ \cite{elength} gives the propagation range of the shear wave. A wave is well-defined only if the wavelength is smaller than the propagation range. This corresponds to $k>\frac{1}{c\tau}$, or approximately to $k>k_g$ with $k_g$ given by (\ref{kgap}).

{Interesting effects related to GMS correspond to the momentum at which the first and second derivative terms in Eq. \eqref{gener3} are of the same order and neither can be neglected, as discussed later in the paper. This corresponds to the breakdown of hydrodynamics intended as a perturbative expansion in spatial and time gradients.} 

Recently, in Ref.\cite{prl}, detailed evidence for GMS was presented on the basis of molecular dynamics simulations. 

The GMS is interesting. Indeed, the two commonly discussed types of dispersion relations are either gapless as for photons and phonons, $E=p$ ($c=1$), or have the energy gap for massive particles, $E=\sqrt{p^2+m^2}$, where the gap is along the Y-axis. On the other hand, (\ref{omega}) implies that the gap is in {\it momentum} space and along the X-axis, similar to the hypothesized tachyon particles with imaginary mass \cite{tachyons}. 

\begin{figure}
\begin{center}
{\scalebox{0.45}{\includegraphics{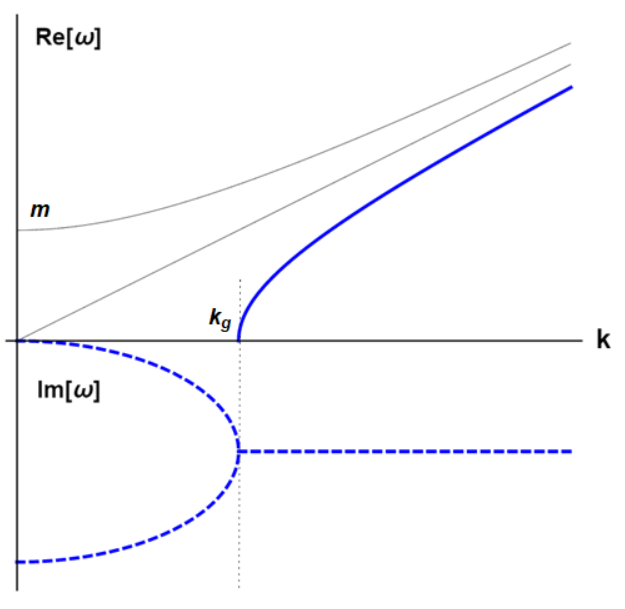}}}
\end{center}
\caption{Possible dispersion relations $\omega(k)$. Top curve shows the dispersion relation for a particle with mass $m$. Middle curve shows gapless dispersion relation for a massless particle (photon) or a phonon in solids. Bottom curve shows the dispersion relation (\ref{omega}) with the gap in $k$-space, $k_g$, illustrating the results of Ref. \cite{prl}. The $k$-gap case displays a non-trivial imaginary part (dashed lines) and the presence of a non-hydrodynamic mode with damping $\mathrm{Im}(\omega)=-1/\tau$ \cite{Baggioli:2018nnp}. Below $k_g$, the dispersion relation of the lowest hydrodynamic mode is purely diffusive.}
\label{3}
\end{figure}

Fig.\ref{3} illustrates the different dispersion relations including the dispersion relation with the $k$-gap. The $k$-gap case displays a non-trivial imaginary part and the presence of a non-hydrodynamic mode with damping $\mathrm{Im}(\omega)=-1/\tau$ \cite{Baggioli:2018nnp}. For small frequency and momentum, the dispersion relation of the lowest mode is purely diffusive and hydrodynamic.

How large can the gap in (\ref{kgap}) get? In condensed matter systems and liquids in particular, $k_g$ is limited by the UV scale: the interatomic separation $a$ and corresponding $k$ point comparable to $\frac{1}{a}$. This can be seen in (\ref{kgap}) by using the shortest value of $\tau$ comparable to the Debye vibration period $\tau_{\rm D}$. Using $\tau=\tau_{\rm D}$ in (\ref{kgap}) gives the maximal value of $k_g$ as $\frac{1}{2a}$, where we used $a=c\tau_{\rm D}$. This corresponds to the shortest wavelength in the system to be $2a$, as expected on general grounds. In this picture, the size of the $k$-gap must be smaller than the UV cutoff of the theory to have a well defined elastic regime where shear waves propagate.

We note that $\tau$ in liquids depends on pressure and temperature. Hence the condition $\tau=\tau_{\rm D}$ giving the maximal $k_g$ gives a well-defined line on the phase diagram. This line is the Frenkel line (FL) separating the combined oscillatory and diffusive molecular motion from purely diffusive motion \cite{ropp,fr1,fr2}. The FL corresponds to qualitative changes of system properties in liquids and supercritical fluids and is ultimately related to the UV cutoff. Close to the FL, where the $k$-gap is maximal, liquids acquire an interesting universality of properties: for example, the liquid viscosity becomes minimal and represents a universal quantum property governed by fundamental physical constants only \cite{minimal} which seems at work even in exotic states of matter such as the quark gluon plasma \cite{Baggioli:2020lcf}.  

Differently from liquids, $k_g$ increases without bound in scale-free field theories discussed below due to the absence of an UV regulator. In these theories, the UV cutoff and the Frenkel line do not exist.

\section{Two-field Lagrangian, its solutions and properties}\label{sectwofields}

\subsection{Two-field Lagrangian}

An important question from the field-theoretical perspective is what Lagrangian gives the spectrum given by Eq. (\ref{omega}) and the associated GMS? The challenge is to represent the viscous term $\propto\tau^{-1}$ in (\ref{gener3}) in the Lagrangian. The viscous energy can be written as the work $W$ done to move the liquid. If $s$ is the strain, $W\propto Fs$, where $F$ is the viscous force $F\propto\eta\frac{ds}{dt}$. Hence, the dissipative term in the Lagrangian should contain the term $s\frac{ds}{dt}$. This can be represented by a scalar field $\phi$, giving the term $L\propto\phi\frac{\partial\phi}{\partial t}$. However, the term $\phi\frac{d\phi}{dt}$ disappears from the Euler-Lagrange equation $\frac{\partial L}{\partial\phi}=\frac{\partial}{\partial t}\frac{\partial L}{\partial\frac{\partial\phi}{\partial t}}+\frac{\partial}{\partial x}\frac{\partial L}{\partial\frac{\partial\phi}{\partial x}}$ because $\frac{\partial L}{\partial\phi}=\frac{\partial}{\partial t}\frac{\partial L}{\partial\frac{\partial\phi}{\partial t}}=\frac{\partial\phi}{\partial t}$. Another way to see this is note that the viscous term is simply a total derivative,  $\phi\frac{d\phi}{dt}\propto\frac{d}{dt}\phi^2$. 

To circumvent this problem, we proposed to operate in terms of {\it two} fields $\phi_1$ and $\phi_2$ \cite{pre,myscirep}. We note that a two-coordinate description of a localised damped harmonic oscillator was discussed earlier \cite{bateman,dekker}. Two fields also emerge in the Keldysh-Schwinger approach to dissipative effects, describing an open system of interest and its environment (bath) \cite{Baggioli:2019jcm}. 

We constructed the dissipative term as the antisymmetric combination of $\phi\frac{d\phi}{dt}$ \cite{pre}, namely as
\begin{equation}
L_d\propto\phi_1\frac{\partial\phi_2}{\partial t}-\phi_2\frac{\partial\phi_1}{\partial t}
\label{ld}
\end{equation}
Taking into consideration this new term, the Lagrangian density involving two scalar fields $\phi_1$ and $\phi_2$ and the dissipative term (\ref{ld}) reads \cite{pre}:
\begin{equation}
L_\phi=\frac{\partial\phi_1}{\partial t}\frac{\partial\phi_2}{\partial t}-c^2\frac{\partial\phi_1}{\partial x}\frac{\partial\phi_2}{\partial x}+\frac{1}{2\tau}\left(\phi_1\frac{\partial\phi_2}{\partial t}-\phi_2\frac{\partial\phi_1}{\partial t}\right)
\label{l1}
\end{equation}

\noindent where we consider only one spatial direction for simplicity. 

The scalar fields $\phi_1,\phi_2$ are real, and we verify the existence of real solutions later in this section. In real space, hermiticity or self-adjointness coincide with the invariance under the transposition operator since complex conjugation acts trivially. Defining the vector $\Phi\equiv (\phi_1,\phi_2)$, we can write down the Lagrangian as:
\begin{equation}
    L_\phi\,=\,\Phi^T\,\mathfrak{L}\,\Phi
\end{equation}
where $T$ indicates the transposition operation. Then, the Lagrangian in Eq. \ref{l1} is not Hermitian in the sense that $\mathfrak{L}^T\neq \mathfrak{L}$. However, we will see later than the Lagrangian \eqref{l1} is $\mathcal{PT}$ symmetric.

Applying the Euler-Lagrange equations to (\ref{l1}) gives two {\it decoupled} equations for $\phi_1$ and $\phi_2$:
\begin{eqnarray}
\begin{split}
c^2\frac{\partial^2\phi_1}{\partial x^2}=\frac{\partial^2\phi_1}{\partial t^2}+\frac{1}{\tau}\frac{\partial\phi_1}{\partial t}\\
c^2\frac{\partial^2\phi_2}{\partial x^2}=\frac{\partial^2\phi_2}{\partial t^2}-\frac{1}{\tau}\frac{\partial\phi_2}{\partial t}
\end{split}
\label{twoeq}
\end{eqnarray}

\noindent where the equation for $\phi_1$ is the same as (\ref{gener3}).

These equations have different solutions depending on whether $k$ is above or below $k_g=\frac{1}{2c\tau}$. For $k>k_g$, the most general form of physically relevant real solutions of \eqref{twoeq} is:

\begin{eqnarray}
\begin{split}
&\phi_1=\bar \phi_1\,e^{-\frac{t}{2\tau}}\cos(\,kx-\omega_R\, t)\\
&\phi_2=\bar \phi_2\,e^{\frac{t}{2\tau}}\,\cos(\,kx-\omega_R\, t+\delta)
\label{twosol}
\end{split}
\end{eqnarray}

\noindent where $\delta$ is the phase shift and where

\begin{equation}
\omega_R=\sqrt{c^2k^2-\frac{1}{4\tau^2}}
\label{gapsol}
\end{equation}

\noindent has the form of (\ref{omega}) predicting the GMS. 

\begin{figure}[t]
    \centering
    \includegraphics[width=0.8 \linewidth]{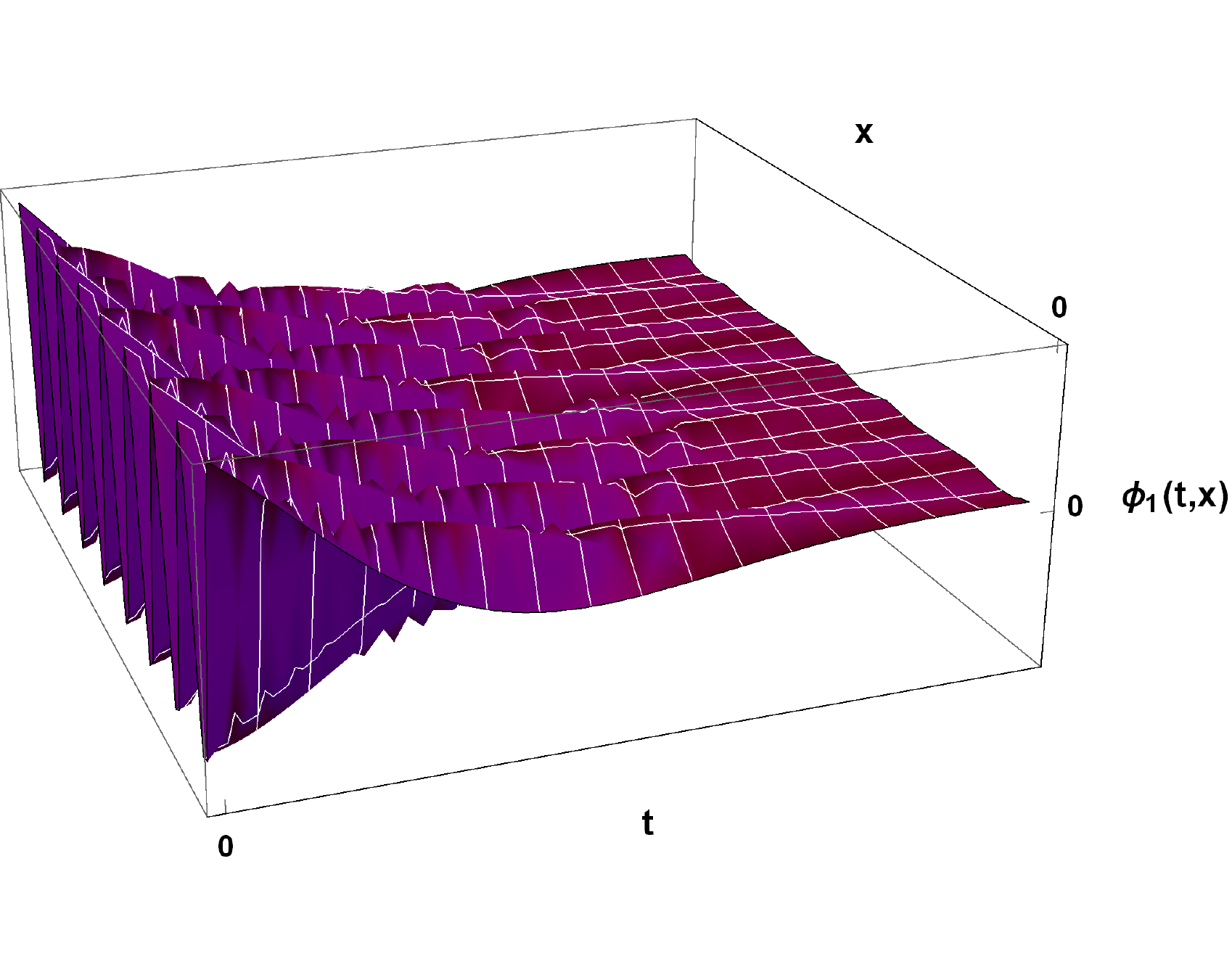}
    \includegraphics[width=0.8\linewidth]{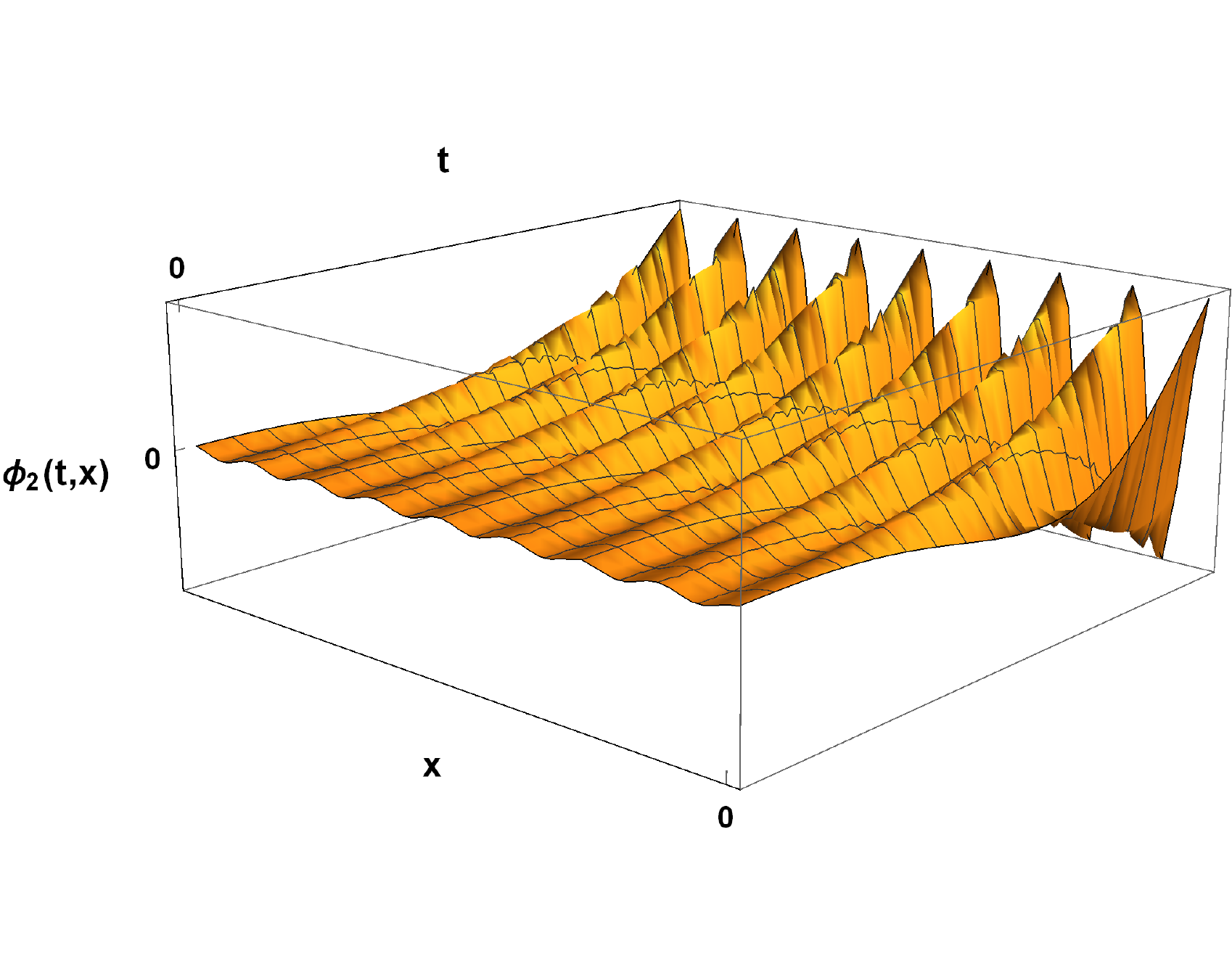}
    \caption{The solutions for $\phi_{1,2}$(t,x) for $k>k_g$. In the time direction there is a wave propagating together with an exponential decay  for the $\phi_1$ field and an exponential increase for the $\phi_2$ field.}
    \label{fig:solsolid}
\end{figure}

The solutions are shown in Fig. \ref{fig:solsolid}.

In the fluid regime\footnote{By ''fluid regime'' we mean the range below $k=k_g$ where no propagating waves appear. Note that the ``fluid regime'' does not correspond to the hydrodynamic regime \textit{stricto sensu}. The second regime needs both the frequency and momentum to be small and accounts for a single diffusive mode in the spectrum, as compared to two modes found at $k<k_g$. In this second definition, hydrodynamics, intended as a perturbative EFT in small gradients, is different from fluid dynamics or fluid mechanics \cite{landau1}. Indeed, it can be applied also to solid systems and crystals \cite{PhysRevA.6.2401,Ammon:2020xyv}.}, $k<k_g$, where no transverse modes propagate, the real solutions are:

\begin{align}
&\phi_1=\bar\phi_1e^{- \alpha_1 t}\cos (kx)\,,\quad \alpha_1=\frac{1}{2\tau}\pm\sqrt{\frac{1}{4 \tau^2}-c^2k^2}\\ 
& \phi_2=\bar\phi_2e^{- \alpha_2 t}\cos (kx+\delta)\,,\quad \alpha_2=-\frac{1}{2\tau}\pm\sqrt{\frac{1}{4 \tau^2}-c^2k^2}
\label{twosol1}    
\end{align}

\begin{figure}
    \centering
    \includegraphics[width=0.8 \linewidth]{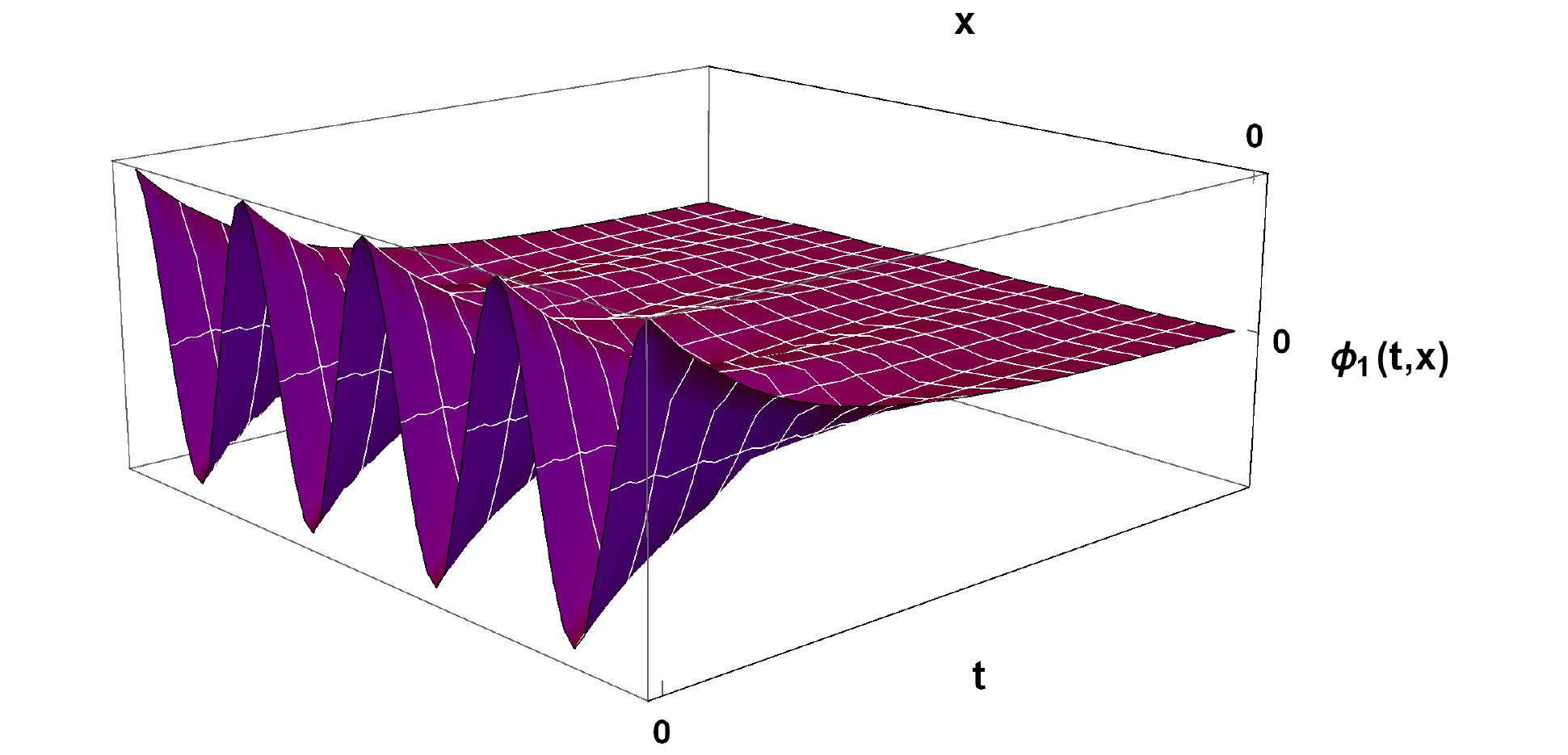}
    \vspace{0.2cm}
    \includegraphics[width=0.8\linewidth]{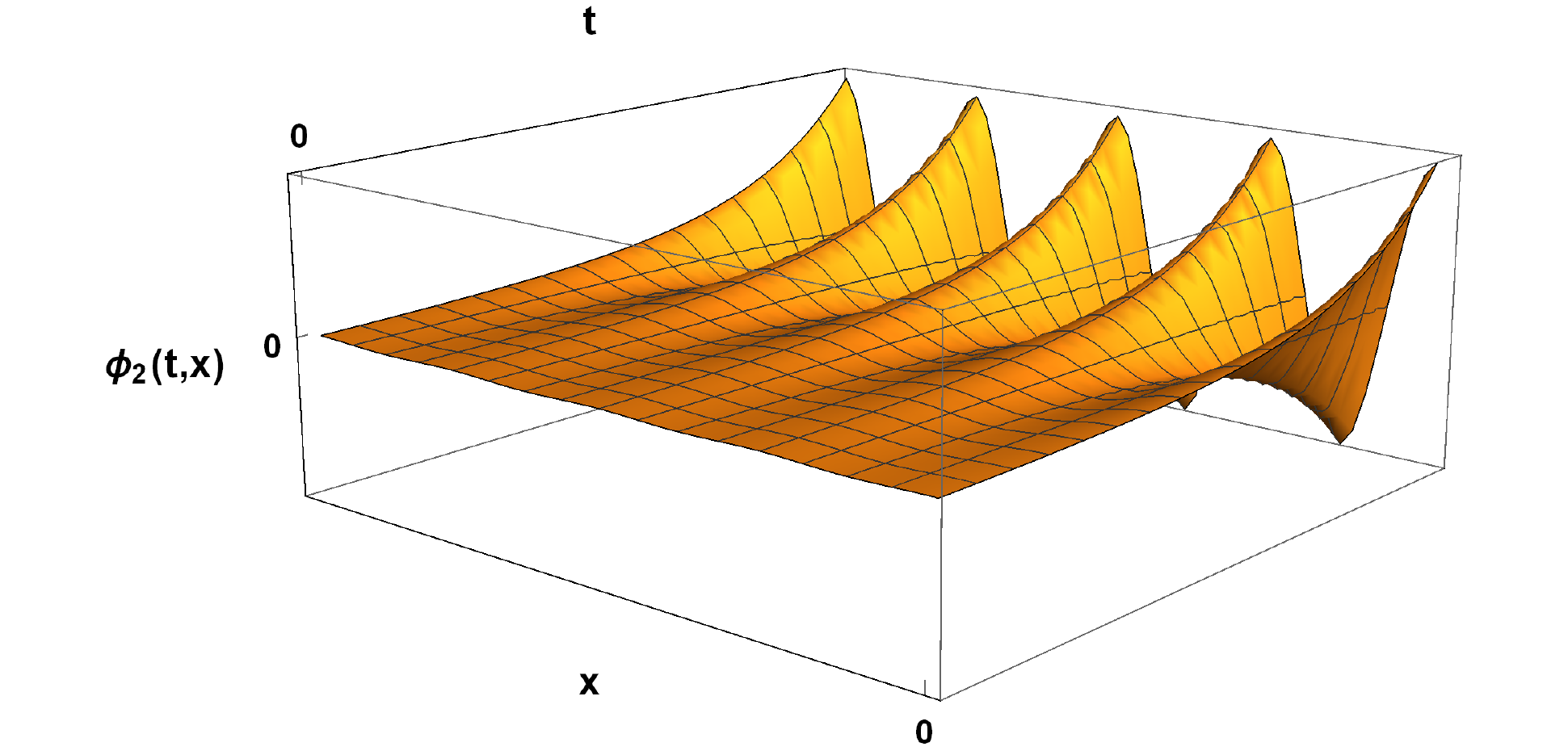}
    \caption{The solutions for $\phi_{1,2}$(t,x) for $k<k_g$. In the time direction there is no wave propagating but a simple exponential decay for the $\phi_1$ field and an exponential increase for the $\phi_2$ field.}
    \label{fig:solhydro}
\end{figure}

These solutions are not periodic in time as illustrated in Fig.\ref{fig:solhydro} but respectively decay and grow exponentially.

The time dependence of $\phi_1$ and $\phi_2$ in (\ref{twosol}) can be interpreted as energy exchange between waves $\phi_1$ and $\phi_2$: $\phi_1$ and $\phi_2$ appreciably decrease and grow over time $\tau$, respectively. This process is similar to the phonon scattering in crystals due to defects or anharmonicity where a plane-wave phonon ($\phi_1$) decays into other phonons (represented by $\phi_2$) and acquires a finite lifetime $\tau$ as a result. After the next time interval $\tau$, the newly created phonon $\phi_2$ decays itself, transferring the energy to other phonons, and the process repeats. The time scale over which we consider and describe the dissipation process in (\ref{twosol}) is $\tau$ because the phonon with the $k$-gap dissipates after time comparable to $\tau$ (\ref{distau}).

This energy exchange can be represented by a generic approach to dissipation where the Lagrangian is written as \cite{endlich}:
\begin{equation}
L[1,2]=L[1]+L[2]+L_{\rm int}[1,2]    
\label{general}    
\end{equation}
The interaction term $L[1,2]$ in (\ref{general}) represents dissipation as the energy transfer from the degrees of freedom ''$1$'' to the degrees of freedom ''$2$''. We keep track of the degrees of freedom ''$1$'' related to dissipation but not the degrees of freedom ''$2$'', either because they are not of interest or are too complicated to account for. $L_{\rm int}[1,2]$ couples the two sectors and represent the energy exchange between them.

In the language of theories studying the open systems and non-Hermitian Hamiltonians (see next section), $\phi_1$ and $\phi_2$ in (\ref{twosol}) are analogous to the ``loss'' and ``gain'' subsystems of a composite system  \cite{benderbook}, although in our case these subsystems are propagating waves suited for field-theoretical description rather than localised oscillators discussed earlier.

The parameters $\tau$ and $k$ in the dispersion relation (\ref{gapsol}) are subject to UV cutoffs in condensed matter phases, including in liquids where GMS emerge as discussed in the previous section. The high-temperature limit of $\tau$ is given by the shortest time scale in the system on the order of Debye vibration period $\tau_{\rm D}$. When $\tau\rightarrow\tau_{\rm D}$, $k_g=\frac{1}{2c\tau}\rightarrow\frac{1}{2c\tau_{\rm D}}\approx\frac{1}{a}$, or $k_{\rm D}$, where $a$ is the inter-particle separation and $k_{\rm D}$ is Debye wavevector. Therefore, the limits of $k$ and $\tau$ at the UV cutoff are 
\begin{equation}
\tau\rightarrow\tau_{\rm D}\,,\quad \quad 
k\rightarrow k_{\rm D}
\label{uv}
\end{equation}

The limits (\ref{uv}) apply to the field theory describing condensed matter phases with a well-defined UV regulator, e.g. lattice spacing. More generally, the field theory discussed here may involve UV cutoffs of different nature depending on the physical object it describes.

\subsection{An alternative formulation of the two-field Lagrangian}

In this section, we further elaborate on the origin of the dissipative Lagrangian discussed in the previous section. This will enable us to make a correspondence with the effective Lagrangian emerging in the Keldysh-Schwinger formalism discussed in the next section. 

The first two cross terms in our initial Lagrangian \eqref{l1} follow from the standard scalar field theory
\begin{equation}
    L_\zeta=\frac{1}{2}\left[\left(\frac{\partial\zeta_1}{\partial t}\right)^2-c^2\left(\frac{\partial\zeta_1}{\partial x}\right)^2+\left(\frac{\partial\zeta_2}{\partial t}\right)^2-c^2\left(\frac{\partial\zeta_2}{\partial x}\right)^2\right]
\end{equation}
using the standard transformation:
\begin{eqnarray}
\begin{split}
\phi_1=\frac{1}{\sqrt{2}}(\zeta_1+i\,\zeta_2)\\
\phi_2=\frac{1}{\sqrt{2}}(\zeta_1-i\,\zeta_2)
\end{split}
\label{trans}
\end{eqnarray}
In terms of the fields $\zeta_1$ and $\zeta_2$, the Lagrangian (\ref{l1}) reads
\begin{eqnarray}
\begin{split}
&L_\zeta=\frac{1}{2}\left[\left(\frac{\partial\zeta_1}{\partial t}\right)^2-c^2\left(\frac{\partial\zeta_1}{\partial x}\right)^2+\left(\frac{\partial\zeta_2}{\partial t}\right)^2-c^2\left(\frac{\partial\zeta_2}{\partial x}\right)^2\right]+\\
&\frac{i}{2\tau}\left(\zeta_2\frac{\partial\zeta_1}{\partial t}-\zeta_1\frac{\partial\zeta_2}{\partial t}\right)
\end{split}
\label{l12}
\end{eqnarray}

We note that in terms of $\zeta$ fields, the non-hermiticity of Lagrangian \eqref{l12} has a more standard meaning of $L_\zeta^\dagger \neq L_\zeta$.

Applying the Euler-Lagrange equations to (\ref{l11}) gives two {\it coupled} equations for $\zeta_1$ and $\zeta_2$ as
\begin{eqnarray}
\begin{split}
c^2\frac{\partial^2\zeta_1}{\partial x^2}=\frac{\partial^2\zeta_1}{\partial t^2}+\frac{i}{\tau}\frac{\partial\zeta_2}{\partial t}\\
c^2\frac{\partial^2\zeta_2}{\partial x^2}=\frac{\partial^2\zeta_2}{\partial t^2}-\frac{i}{\tau}\frac{\partial\zeta_1}{\partial t}
\end{split}
\label{twoeq1}
\end{eqnarray}
These equations can be de-coupled by using the same transformation (\ref{trans}): using (\ref{trans}) in (\ref{twoeq1}) gives the system of two equations for $\phi_1$ and $\phi_2$. Adding and subtracting these equations gives (\ref{twoeq}). 

The solutions of (\ref{l12}) and (\ref{twoeq1}) are generally complex. Representing these fields in the Lagrangian can be done using complex field conjugates \cite{alexandre1,alexandre2}. However, this introduces an ambiguity in the equations of motion derived by applying the Euler-Lagrange equations to the Lagrangian. The ambiguity can be removed by selecting the solutions related by ${\mathcal PT}$ symmetry \cite{alexandre1,alexandre2} (see next section).   

We note that taking real solutions $\phi_1$ and $\phi_2$ in (\ref{twosol}), together with (\ref{trans}), implies that $\zeta_1$ is real and $\zeta_2$ is purely imaginary. In order to have a clearer formulation of our field theory, we continue our Lagrangian formulation in terms of real fields, similarly to (\ref{twosol}) and define new real fields $\psi_1$ and $\psi_2$ as $\zeta_1=\psi_1$ and $\zeta_2=-i\psi_2$. Then, (\ref{trans}) becomes
\begin{eqnarray}
\begin{split}
\phi_1=\frac{1}{\sqrt{2}}(\psi_1+\psi_2)\\
\phi_2=\frac{1}{\sqrt{2}}(\psi_1-\psi_2)
\end{split}
\label{trans1}
\end{eqnarray}
where all the fields involved in this transformation are real valued.

In the next section, we will see that this transformation is the same as the ``Keldysh rotation'' used in the Keldysh-Schwinger formalism.

In terms of the fields $\psi$, the Lagrangian (\ref{l1})  reads
\begin{eqnarray}
\begin{split}
&L_\psi=\frac{1}{2}\left[\left(\frac{\partial\psi_1}{\partial t}\right)^2-c^2\left(\frac{\partial\psi_1}{\partial x}\right)^2-\left(\frac{\partial\psi_2}{\partial t}\right)^2+c^2\left(\frac{\partial\psi_2}{\partial x}\right)^2\right]+\\
&+\frac{1}{2\tau}\left(\psi_2\frac{\partial\psi_1}{\partial t}-\psi_1\frac{\partial\psi_2}{\partial t}\right)
\end{split}
\label{l11}
\end{eqnarray}

The Hamiltonian based on (\ref{l1}) or (\ref{l11}) is non-Hermitian due to the presence of the anti-symmetric last term $\sim 1/\tau$, but it is $\mathcal{PT}$ invariant (see Ref. \cite{Bender:2004sv} for a similar case). As compared to the free-field part of (\ref{l12}), the free term for $\psi_2$ in (\ref{l11}), $-\left(\frac{\partial\psi_2}{\partial t}\right)^2+c^2\left(\frac{\partial\psi_2}{\partial x}\right)^2$, has the opposite sign. In the next section, we show that this is related to the result following from Keldysh-Schwinger formalism. The opposite sign is not an issue from the point of view of system's energy because the kinetic matrix is not diagonal due to the coupling $\sim\frac{1}{\tau}$. As explicitly shown in the next section, the system is stable and the Hamiltonian has a well-defined lower bound. 

Let consider the case $\bar\phi_1=\bar\phi_2=1$ and $\delta=0$ in \eqref{twosol}. Using (\ref{twosol}) and (\ref{trans1}), the solutions in terms of $\psi_1$ and $\psi_2$ for $k>k_g$ are
\begin{eqnarray}
\begin{split}
    &\psi_1(t,x)=\frac{1}{\sqrt{2}}{e^{-\frac{t}{2 \tau }} \,\left(e^{t/\tau }+1\right)\, \cos (k x-\omega_R t)}\\
    &\psi_2(t,x)=\frac{1}{\sqrt{2}}{e^{-\frac{t}{2 \tau }} \,\left(e^{t/\tau }-1\right)\, \cos (k x-\omega_R  t)}
\end{split}    
\label{twopsi}
\end{eqnarray}

For $k<k_g$, the solutions are
\begin{eqnarray}
\begin{split}
    &\psi_1(t,x)=\frac{1}{\sqrt{2}}\cos (k x)\,e^{-t\,( \alpha_1+\alpha_2)}\left(e^{\alpha_1 \,t}+e^{\alpha_2 \,t}\right)\\
    &\psi_2(t,x)=\frac{1}{\sqrt{2}}\cos (k x)\,e^{-t\,( \alpha_1+\alpha_2)}\left(e^{\alpha_1 \,t}-e^{\alpha_2\, t}\right)
\end{split}    
\end{eqnarray}
 In the limit $\tau\rightarrow\infty$, $\psi_2=0$ according to (\ref{twopsi}). Hence, in the absence of dissipation, we have
\begin{eqnarray}
\begin{split}
&\psi_1(t,x)\,=\,\cos[k(x-ct)]\\
&\psi_2(t,x)\,=\,0
\end{split}
\label{onefield}
\end{eqnarray}
In this case, the Lagrangian (\ref{l11}) becomes the free-field Lagrangian for the single field $\psi_1$. Using $\psi_2=0$ in (\ref{trans1}) also implies that $\phi_1=\phi_2$, and the Lagrangian (\ref{l1}) similarly becomes the free-field Lagrangian for the single field $\Phi\equiv\phi_1=\phi_2$.\color{black}

This is an important point for two reasons. First, it shows that the number of degrees of freedom is halved in the absence of dissipation at $\tau=\infty$. The meaning of the additional degree of freedom will become clear in the Keldysh-Scwhinger formalism in section \ref{six}. Second, the remaining single degree of freedom is a plane wave with the dispersion relation $\omega=ck$. This ensures that the system reduces to the canonical situation in the limit $\tau \rightarrow \infty$.

\subsection{The Lagrangian from Keldysh-Schwinger formalism} \label{six}

In the previous sections, we discussed how describing dissipation
necessitates a two-field Lagrangian. We have earlier noted that the
Keldysh-Schwinger (KS) formalism similarly involves two fields, and that
the Green function operator contains the first time derivative which can
be related to GMS \cite{Baggioli:2019jcm}. Here, we make a stronger and
more specific assertion. We show that two new important features of our dissipative Lagrangian (\ref{l11}) and emerging GMS appear in the KS formalism: (a) the new dissipative term of the form (\ref{ld}),
$\left(\psi_2\frac{\partial\psi_1}{\partial
t}-\psi_1\frac{\partial\psi_2}{\partial t}\right)$ and (b) the opposite
sign of the free-field term of $\psi_2$ in $L_\psi$ (third and fourth
terms in (\ref{l11})).

The description of non-equilibrium effective field theories involves the doubling of the degrees of freedom \cite{Glorioso:2018wxw}. Let us consider the simple case of one scalar field, $\psi_1$. The action of a dissipative system depends on the initial state of the system and, accordingly, on the choice of initial time. The averaging operation is not defined in this case and, as a consequence, the statistical theory can not be formulated. To get around this problem, the KS formalism uses the following approach: consider a copy of our system with the same transition amplitude and the field in the replica system $\psi_2$. Recall that both fields are identical, hence $\langle \psi_1(0) | \psi_2(0 )\rangle = 1 $. Using these two fields, we reverse (invert) time in the second system and close the integration contour at $ t = \infty $. Then we can write the system's path integral as
\begin{equation}
   e^{W(\zeta_1,\zeta_2)}\,=\,\int D\psi_1 D\psi_2
\,e^{i\,\int\limits^t_{\infty}\mathrm{d}t\left(\mathcal{L}(\psi_1;\xi_1)\,-\,\mathcal{L}(\psi_2;\xi_2)\right)\,+\,i\,\mathcal{L}_{1,2}(\infty)}
\end{equation}

\noindent where $\xi_{1,2}$ are source terms associated with fields $\psi_{1,2}$. 

\begin{figure}[h!]
     \centering
     \includegraphics[width=0.6\linewidth]{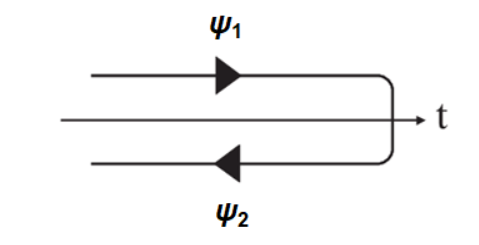}
     \caption{The opposite sign in front of field $\psi_2$ follows from closing the integration contour in the KS formalism.}
     \label{figsc}
\end{figure}

Importantly, the term related to the copy $\psi_2$ has the opposite sign, meaning that it propagates backward
along the Keldysh-Schwinger contour \cite{kamenev,Glorioso:2018wxw}. Therefore, we can interpret the field $\psi_2$ as the additional degree
of freedom required to describe an open system and whose dynamics is reversed with respect to the arrow of time (see Fig. \ref{figsc}).

The KS formalism involves introducing $r$,$a$ variables (retarded/advanced), defined as:

\begin{equation}
\psi_r\,=\,\frac{1}{\sqrt{2}}\,\left(\psi_1\,+\,\psi_2\right)\,,\quad
\psi_a\,=\,\frac{1}{\sqrt{2}}\,\left(\psi_1\,-\,\psi_2\right)
\end{equation}

\noindent where variable $a$ is related to the real field dynamics,
and variable $r$ related to dissipation and quantum fluctuations.
This is the same transformation as (\ref{trans1}) which we used earlier
to find the relation between fields $\psi$ and $\phi$.

By comparing the KS tranformation with (\ref{trans1}), we see that
$\phi_2$ plays the role of $\psi_a$ which corresponds to dissipation. As
we will later find, $\langle\phi_2\phi_2 \rangle=0$. This implies
unitarity of our field theory, ensured by the $\mathcal{PT}$ invariance
of the action \cite{Mannheim:2009zj}.

We now make this more precise and elaborate on the details. Let us consider
an out-of-equilibrium system represented by a continuous field whose
energy dissipates with time. We assume that at $t=0$, the system is out
of equilibrium and is in the state $ | \Phi_0 \rangle $. It evolves to
its final state, $\langle \Phi_{\infty} |$ with energy is
$E_{\infty}=0$. In order to describe the system's non-equilibrium
dynamics, we use the KS technique \cite{kamenev_2011}. We first write
the transition probability from the initial state to the  final one as
follows:
\begin{align}
&\langle \Phi_{\infty}, \infty |\Phi_0, 0 \rangle = \langle
\Phi_{\infty}|\mathcal{\hat U}_{\infty} |\Phi_0 \rangle = \\ \nonumber
&\int \mathfrak{D}\Phi_{\infty}\mathfrak{D}\Phi
\exp
\left[i\int\limits^{\infty}_{0}\mathrm{d}t\,\left(\mathcal{L}(\Phi)
+i\int\limits_V\dfr{\mathrm{d}V}{V\tau}\bar{\Phi}\partial_t\Phi\right)\right]
\end{align}

\noindent where $\bar{\Phi}$ is the field conjugated to $\Phi$, $\tau$ is the characteristic time and the Lagrangian is given by
\begin{gather*}
\mathcal{L}(\Phi
)=\dfr1V\int\limits_{V}\mathrm{d}V\left(\dot \Phi
^2-c^2(\nabla\Phi)^2\right)
\end{gather*}
For details of this derivation we refer the reader to Ref. \cite{Baggioli:2019jcm}.

As discussed earlier, this probability depends on the initial state of
the system and, accordingly, on the choice of the initial time. The
averaging operation is not defined in this case and, as a consequence,
the statistical theory can not be formulated. In order to get around
this problem, the KS approach \cite{kamenev_2011} introduces a copy of
the system with the same transition probability. We denote the field in
the initial system as $\Phi^+$ and in the copy as $\Phi^-$. This is the
same field, hence $\langle \Phi_0^-, 0 | \Phi_0^+, 0 \rangle = 1 $.
Using the two fields, we close the integration contour in point
$t=\infty$ (see Fig. \ref{figsc}) and write\begin{gather*}
1\equiv \langle \Phi_0^- , 0|\Phi_0^+ , 0 \rangle=
\int \mathfrak{D}\Phi_{\infty}\langle \Phi_0^- , 0| \Phi_{\infty},
\infty \rangle
%\langle \phi_{eq}| \phi_{eq}\rangle
\langle \Phi_{\infty}, \infty |\Phi_0^+ , 0 \rangle
\end{gather*}

After the Wick rotation $t\to it$, we obtain
\begin{align}
&\langle\Phi_0^-, 0 | \Phi_0^+, 0\rangle =
\mathcal{N}'\int \mathfrak{D}\Phi^+ \mathfrak{D}\Phi^-\exp
\Big[-\int\limits^{\infty}_{0}\mathrm{d}t\,\Big(\mathcal{L}(\Phi^+)-
\nonumber\\
&
-\int\limits_V\dfr{\mathrm{d}V}{V\tau}\bar{\Phi^+}\partial_t\Phi^+-\mathcal{L}(\Phi^-)
+\int\limits_V\dfr{\mathrm{d}V}{V\tau}\bar{\Phi^-}\partial_t\Phi^-\Big) \Big]
\end{align}
Therefore, the effective Lagrangian of the theory is given by
\begin{eqnarray}
\begin{split}
&\mathcal{L}'=\left(\dfr{\partial \Phi^+}{\partial
t}\right)^2-c^2\left(\dfr{\partial \Phi^+}{\partial
x}\right)^2-\left(\dfr{\partial \Phi^-}{\partial
t}\right)^2+c^2\left(\dfr{\partial \Phi^-}{\partial x}\right)^2-\\
& -\dfr1{\tau}\left(\bar{\Phi}^+\dfr{\partial \Phi^+}{\partial
t}-\bar{\Phi}^-\dfr{\partial \Phi^-}{\partial t}\right)
\end{split}
\label{lagks0}
\end{eqnarray}
We now perform Keldysh rotation:
\begin{equation}
    \phi_1=\frac{1}{\sqrt{2}}\,(\Phi^++\Phi^-)\,,\,\quad
\phi_2=\frac{1}{\sqrt{2}}\,(\Phi^+-\Phi^-)\,\,,
\end{equation}

\noindent which is the same transformation we used earlier
(\ref{trans1}), and find

\begin{equation}
\mathcal{L}'=\dfr{\partial \bar\phi_1}{\partial t}\dfr{\partial
\phi_2}{\partial t}-c^2\dfr{\partial\bar\phi_1}{\partial x}\dfr{\partial
\phi_2}{\partial x}+
\dfr1{\tau}\left(\bar\phi_1\dfr{\partial \phi_2}{\partial t}-\dfr{\partial
\bar\phi_1}{\partial t}\phi_2\right)
\label{lagks}
\end{equation}

\noindent where we use the commutation relation for bosonic operators
$\left[\Phi^a,\Phi^b\right]=0$ and note that $\dfr{\partial
\phi_1}{\partial t}\,\bar\phi_2=-\left(\overline{\dfr{\partial \phi_1}{\partial
t}}\right)\phi_2=-\dfr{\partial \bar\phi_1}{\partial
t}\phi_2$.

Setting $\bar\phi_1=\phi_1$, we observe that (\ref{lagks}) coincides with
the dissipative Lagrangian we started with in (\ref{l1}) (up to a factor
of 2 in front of $\tau$).

It is remarkable that the dissipative Lagrangian (\ref{l1}) and,
therefore, the gapped momentum states effect appear to be related to a
mature technique such as Keldysh-Schwinger formalism. We note an
important caveat of this relation. The KS formalism does not contain the 
relaxation time $\tau$. Instead, the timescale in the KS approach is
set by the Planck constant. We have introduced $\tau$ as the relaxation time
of the system on top of the standard KS formulation
\cite{Baggioli:2019jcm}, similarly to how $\tau$ is introduced in the
Maxwell-Frenkel interpolation in (\ref{a1})-(\ref{gener3}), where it is
assigned the meaning of the time between molecular rearrangement in the
liquids \cite{frenkel}. In this sense, Lagrangian (\ref{lagks0}) is not
a derivation of our dissipative Lagrangian (\ref{l11}) from the KS
formalism. However, it backs up two new and important features of our
dissipative Lagrangian (\ref{l11}). First, it contains the term of the
form $\propto\phi_1\frac{\partial\phi_2}{\partial
t}-\phi_2\frac{\partial\phi_1}{\partial t}$. This is the same term
(\ref{ld}) featuring in our Lagrangians (\ref{l1}) and (\ref{l11}) which
is required to obtain GMS and which enters with the time scale set by
$\tau$. Second, the third and fourth terms describing the free-field
contribution of the second field enter (\ref{lagks0}) with the sign
opposite to that of the first field: $-\left(\dfr{\partial
\Phi^-}{\partial t}\right)^2+c^2\left(\dfr{\partial \Phi^-}{\partial
x}\right)^2$. This is the same as in our dissipative Lagrangian
(\ref{l11}).

\subsection{The Hamiltonian}

The Hamiltonian of our composite system consisting of fields $\phi_1$ and $\phi_2$ is
\begin{equation}
    H_\phi=\pi_1\frac{\partial\phi_1}{\partial t}+\pi_2\frac{\partial\phi_2}{\partial t}-L_\phi
\end{equation}

\noindent where $L_\phi$ is given in (\ref{l1}) and where the conjugate momenta are
\begin{equation}
    \pi_1=\frac{\partial\phi_2}{\partial t}-\frac{\phi_2}{2\tau}\,,\quad \pi_2=\frac{\partial\phi_1}{\partial t}+\frac{\phi_1}{2\tau}\,.
\end{equation}
This gives 
\begin{equation}
H_\phi=\frac{\partial\phi_1}{\partial t}\frac{\partial\phi_2}{\partial t}+c^2\frac{\partial\phi_1}{\partial x}\frac{\partial\phi_2}{\partial x}
\label{hamilt2}
\end{equation}
The terms with $\tau$ re-appear in the Hamiltonian once the Hamiltonian is written in terms fields and momenta, as discussed in the next section.

We now compute the energy of our system directly using the solutions derived in the previous sections. The results are obviously independent of the choice of variables such as $\phi$ or $\psi$.

\begin{figure}
    \centering
    \includegraphics[width=1\linewidth]{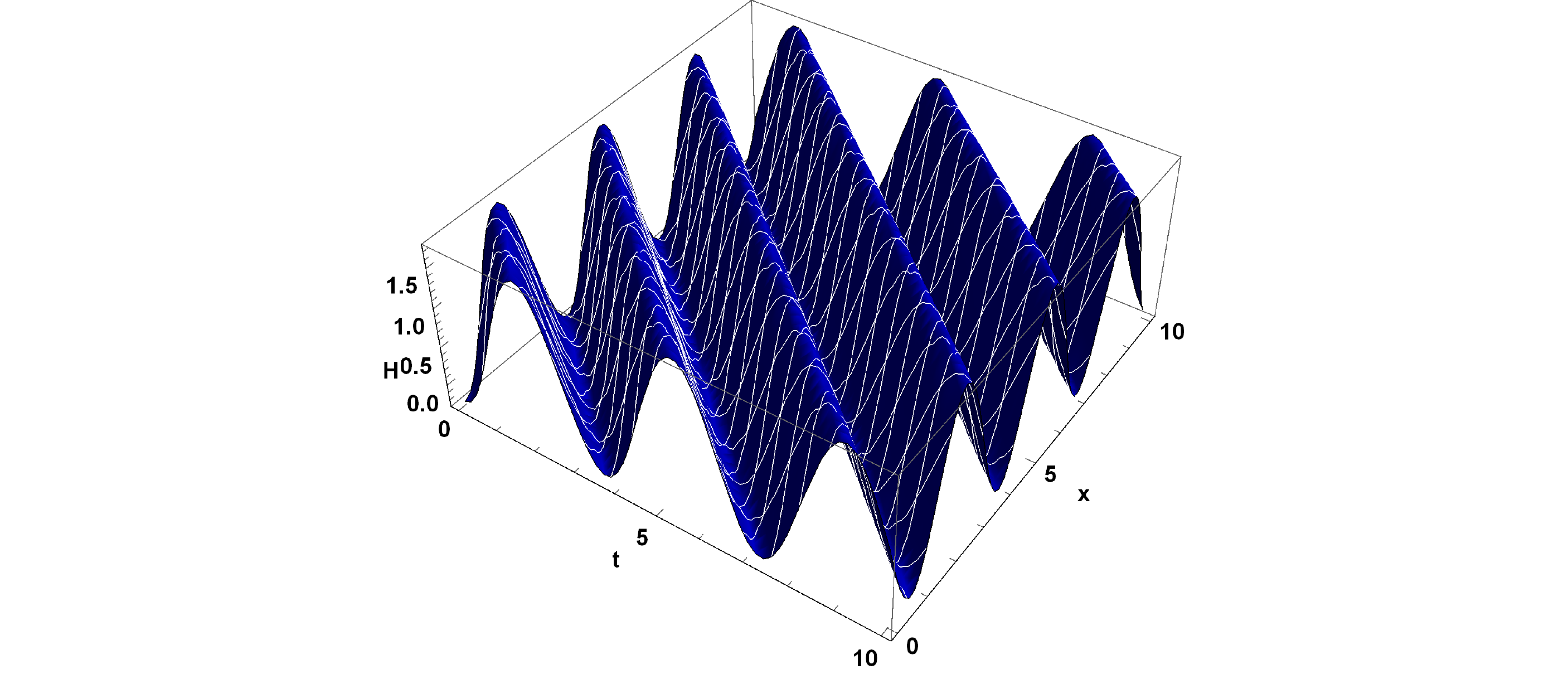}
    \vspace{0.2cm}
    \includegraphics[width=1\linewidth]{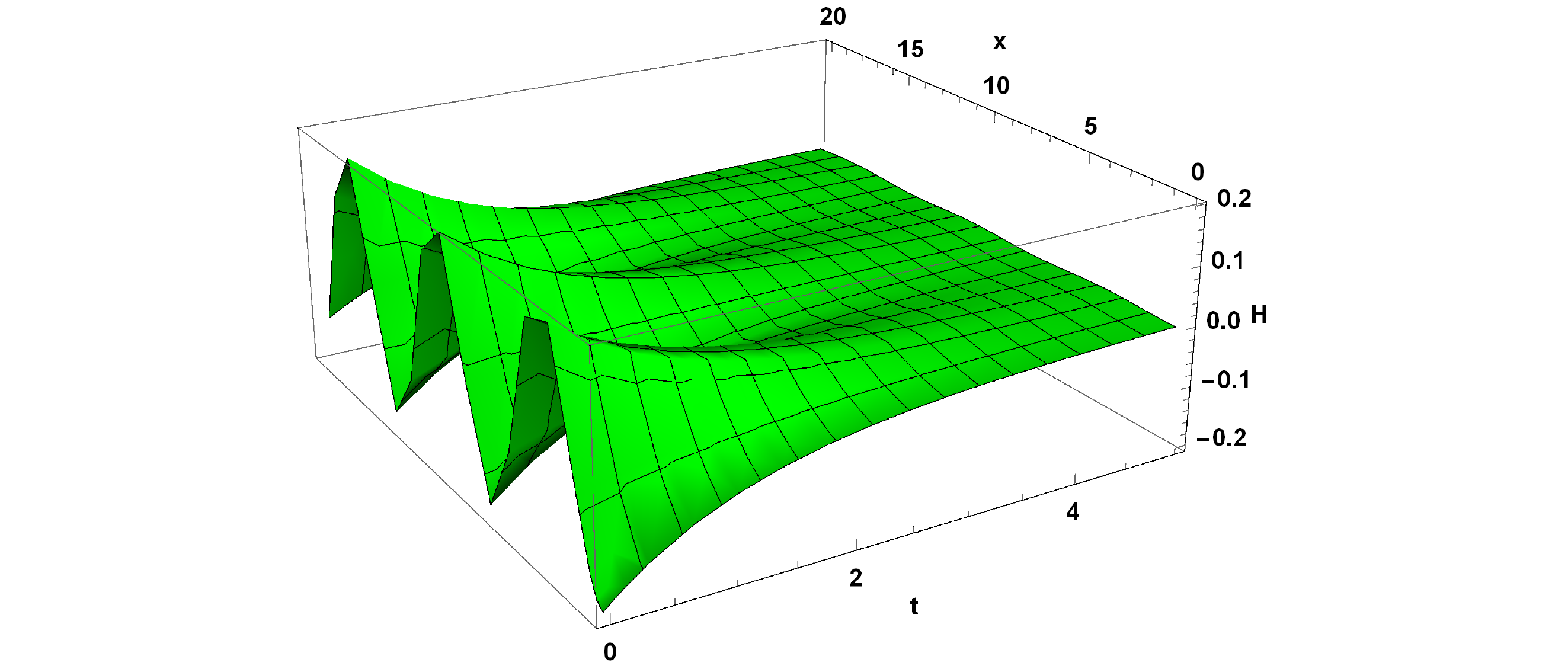}
    \caption{The Hamiltonian $H(t,x)$ for $k>k_g$ (top) and $k<k_g$ (bottom).}
    \label{fig:ham}
\end{figure}

Using the solutions for $\phi$ in (\ref{twosol}), the Hamiltonian for $k>k_g$ above the $k$-gap is
\begin{equation}
H\,=\,\omega_R^2-c^2 k^2 \cos \left(2\,(\omega_r t - k x)\right)
\label{hs1}    
\end{equation}
Below the $k$-gap, we have:
\begin{equation}
H\,=\,-c^2 k^2 \cos (2 k x)\, e^{-2\omega_R\,t}
\label{hs2}
\end{equation}
At $k=k_g$, $\omega_R=0$, and the two results coincide:
\begin{equation}
    H_{k_g}\,=\,-c^2\,k^2\,\cos(2\,k\,x)
\end{equation}
The Hamiltonian is displayed in Fig. \ref{fig:ham} in both regimes. The Hamiltonian oscillates both in time and space for $k>k_g$ and decays with time for $k<k_g$. 

The lower bound of (\ref{hs1}) is $H_l=\omega_R^2-c^2k^2$, or $-\frac{1}{4\tau^2}$, according to (\ref{gapsol}). The lower limit of (\ref{hs2}) is $H_l=-c^2 k^2$. Given that $k<k_g$ in this regime, $H_l$ of (\ref{hs2}) is $-c^2k_g^2$. Combined with $k_g=\frac{1}{2c\tau}$ from (\ref{gapsol}), the lower bound of (\ref{hs2}) is $H_l=-\frac{1}{4\tau^2}$, the same as the lower bound of (\ref{hs1}). Therefore, $H_l$ has a finite value in both regimes for a finite $\tau$. We recall that the UV cutoff for $\tau$ in condensed matter systems is given by $\tau_{\rm D}$ in (\ref{uv}). The limit $\tau\rightarrow 0$ corresponds to the infinite gap $k_g\propto\frac{1}{\tau}$ and, therefore, non-propagating waves which our Lagrangian formulation is not designed to describe.

We now perform time and space average of the Hamiltonians (\ref{hs1}) and (\ref{hs2}) and find
\begin{equation}
\langle H \rangle=\omega_R^2>0 \quad \text{for}\quad k>k_g
\label{hav1}
\end{equation}

\noindent in agreement with the earlier calculation \cite{pre}, and
\begin{equation}
\langle H \rangle=0\quad \text{for}\quad k \leq k_g
\label{hav2}
\end{equation}

(\ref{hav2}) is consistent with the fact that there are no propagating waves below the $k$-gap.

To summarise, we find that the Hamiltonian of the composite gain-loss system is stationary in the propagating wave regime $k>k_g$. In this regime, the energy has a lower bound and a positive average value. In the non-propagating regime $k<k_g$, the system energy similarly has a lower bound and zero average energy as expected.

 \section{Non-Hermiticity and ${\mathcal PT}$ symmetry}\label{secherm}
 
As mentioned earlier, the Lagrangian of our theory, (\ref{l1}) or (\ref{l11}), is not Hermitian. However, we will see that both the Lagrangian and Hamiltonian in our theory are $\mathcal{PT}$ symmetric. We first recall how non-Hermiticity arises in theoretical approaches to dissipation. 

The effect of dissipation can be generally represented by a complex energy spectrum, with imaginary term setting the lifetime of the state (see, e.g., \cite{sudarshan,sud1,rotter2015review,rotter1}). This is similarly discussed in the context of resonances (see Ref. \cite{Feinberg:2010xw} for a recent discussion and a review) in which the complex energy plane is considered. Resonances are derived from complex poles of the form:
\begin{equation}
    \varepsilon _n\,=\,\epsilon_n\,-\,i\,\Gamma_n \label{cc}
\end{equation}

\noindent which necessitates a non-Hermitian model. The width of the resonances or, equivalently, their lifetime, is due to dissipation.

To illustrate how non-Hermiticity is related to dissipation and a finite relaxation time, it is instructive to consider a simple quantum mechanical system in the Heisenberg picture \cite{Chernodub:2019ggz}. Given a generic operator $\mathcal{O}$, its time dependence is given by:
\begin{equation}
    \mathcal{O}(t)\,=\,e^{i\,H^\dagger\,t}\,\mathcal{O}\,e^{-\,i\,H\,t}
\end{equation}
The dynamics of such operator is:
\begin{equation}
    \frac{d \mathcal{O}(t)}{dt}\,=\,i\,e^{i\,H^\dagger\,t}\,\,\left(H^\dagger\,\mathcal{O}\,-\,\mathcal{O}\,H\right)\,e^{-\,i\,H\,t}
\end{equation}
We observe that a Hermitian Hamiltonian $H^\dagger=H$ implies the conservation of this operator. On the contrary, the non-Hermiticity introduces a finite relaxation time:

\begin{equation}
    \frac{d \mathcal{O}(t)}{dt}\,\neq\,0
\end{equation}

The energy spectrum of a Hermitian Hamiltonian are real, however one of the central points of the discussion of symmetry under parity and time transformations (${\mathcal PT}$ symmetry) is that a Hermiticity can be replaced by a weaker condition: a non-Hermitian but ${\mathcal PT}$-symmetric Hamiltonian may still result in real spectra. This follows from an assertion that ${\mathcal PT}$-symmetric Hamiltonians have secular equations with real coefficients so that some of the eigenvalues can be real depending on parameters \cite{benderbook,bender} \footnote{It was observed that ''\textit{The reality of the spectrum of $H$ implies the presence of an antilinear symmetry (which is not necessarily $\mathcal{PT}$). Moreover, the spectrum of $H$ is real if and only if there is a positive-definite inner-product on the Hilbert space with respect to which $H$ is Hermitian or alternatively there is a pseudo-canonical transformation of the Hilbert space that maps $H$ into a Hermitian operator}'' \cite{Mostafazadeh:2001jk,Mostafazadeh:2001nr,Mostafazadeh:2002id}.}. 

The discussion of the ${\mathcal PT}$ symmetry \cite{benderbook} starts with noting that a realistic physical system is an {\it open} non-isolated system with accompanying flux of probability flowing in or out. Theoretical description of this system is a long-standing problem, both classically and quantum-mechanically \cite{davies1976quantum,rotter2015review}. The proposal to address this problem is to treat an open system as a subsystem and add another, time-reversed, subsystem with the opposite net flux of probability, so that the composite system had no net gain or loss of probability flux and is closed. The composite system exhibits the ${\mathcal PT}$ symmetry, where ${\mathcal T}$ is the time-reversal operator and ${\mathcal P}$ is the generic parity operator that interchanges the two subsystems \cite{benderbook}.

\begin{figure}
    \centering
    \includegraphics[width=0.93 \linewidth]{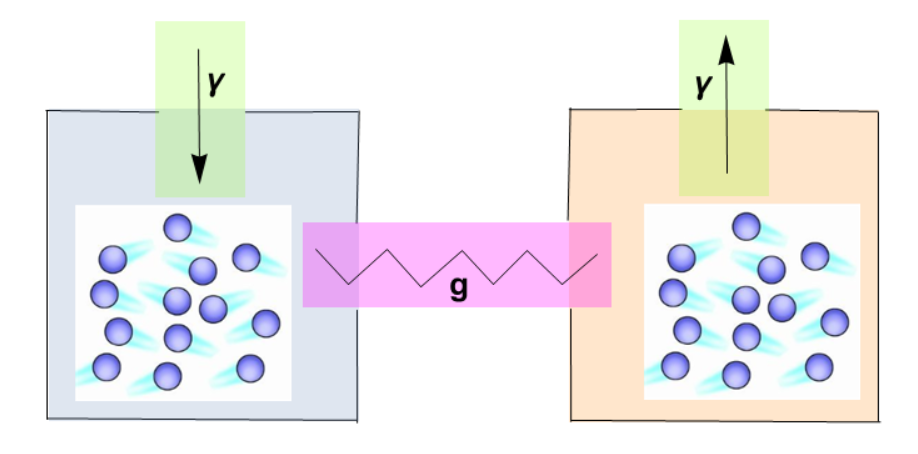}
    \caption{The gain-loss mechanism typical of non-Hermitian systems. One system is dissipating at a rate $\gamma$, while the other is absorbing at the same rate. If the coupling between the two systems $g$ is large enough, ${\mathcal PT}$ symmetry is unbroken and the eigenvalues are real.}
    \label{fig2}
\end{figure}

Although the two subsystems are not stationary and are not in equilibrium separately, the stationary state of the composite system can be achieved by coupling the subsystems. The eigenvalues of the composite system are real, provided the coupling parameter is large enough, corresponding to the stationary state of the system and unbroken ${\mathcal PT}$ symmetry \cite{benderbook}. This process can be illustrated by a system two coupled localised oscillators with gain and loss discussed earlier (see Fig. \ref{fig2}). The equations of motions for two coupled oscillators are $\ddot{x}+\omega^2x+\gamma\dot{x}=gy$ and $\ddot{y}+\omega^2y-\gamma\dot{y}=gx$, where $x$ and $y$ are coordinates, $\gamma$ is the friction coefficient and $g$ is the coupling parameter. The coupling term enters the Hamiltonian as $H_c=-gxy$, and the total Hamiltonian is ${\mathcal PT}$-symmetric. There are three regimes: weak, intermediate and strong coupling corresponding to no real solutions (frequencies), four real solutions and two real solutions, respectively. The state where all solutions are real corresponds to unbroken ${\mathcal PT}$ symmetry, whereas complex solutions correspond to broken symmetry  \cite{benderbook}.

There are several interesting similarities and differences between the above system discussed in the context of ${\mathcal PT}$ symmetry and the system in our theory. First, $\phi_1$ and $\phi_2$ describe fields, rather than localised oscillators and correspond to propagating waves in the solutions. This is required in order to describe the $k$-dependence and gapped momentum states in particular. Second, the $\phi_1$ and $\phi_2$ in (\ref{twosol}) can be viewed as two subsystems with opposite fluxes of probability, similarly to the discussion of ${\mathcal PT}$ symmetry above. Third, the coupling term between $\phi_1$ and $\phi_2$ ($\psi_1$ and $\psi_2$) is different and involves the coupling between one field and the derivative of the other field (see Eqs. (\ref{ld}),(\ref{l1}) and (\ref{l11})) rather than between the fields themselves as in the model used in the above discussion of ${\mathcal PT}$-symmetry.

Let us look at the properties of our system in more detail. The easiest way is to define the doublet:
\begin{equation}
    \Phi\,\equiv\,\begin{pmatrix}
          \phi_1(\vec{x},t) \\
          \phi_2(\vec{x},t)
   \end{pmatrix}
\end{equation}

Parity ${\mathcal P}$ and time reversal ${\mathcal T}$ transformations act on the coordinates as:
\begin{align}
    &\mathcal P\,:\,\quad \vec{x}\,\rightarrow\,-\,\vec{x}\,,\quad \quad  \mathcal T\,:\,\quad t\,\rightarrow\,-\,t
\end{align}

Their action on the field doublet can be written in matrix form as:

\begin{equation}
    \mathcal{T}\,=\,\begin{pmatrix}
          1 & 0\\
          0 &1
   \end{pmatrix}\,,\quad \quad \mathcal{P}\,=\,\begin{pmatrix}
          0 & 1\\
          1 & 0
   \end{pmatrix}\,\quad \longrightarrow \quad \mathcal{PT}\,=\,\begin{pmatrix}
          0 & 1\\
          1 & 0
   \end{pmatrix}
\end{equation}
The latter coincides with the statement that a $\mathcal{PT}$ transformation swaps the source and the sink \cite{benderbook}, \textit{i.e.} $\phi_1 \leftrightarrow \phi_2$.

\begin{figure}
    \centering
    \includegraphics[width=0.9 \linewidth]{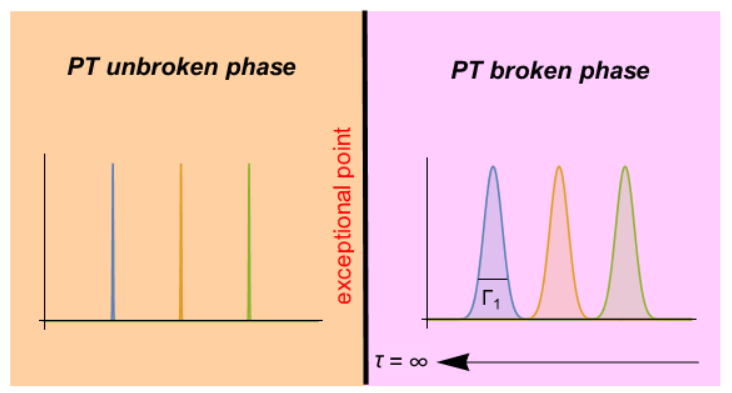}
    \caption{In the $\mathcal{PT}$ unbroken phase all the eigenvalues are real, and the excitation lifetimes are infinite. In the $\mathcal{PT}$-broken phase, finite relaxation times appear. From a mathematical point of view the separation is given by criterion \eqref{crit}. In our case, the exceptional point is at $\tau=\infty$, and we are always in the $\mathcal{PT}$-broken phase where the relaxation time is finite.}
    \label{fig3}
\end{figure}

Given these definitions, we observe that the Lagrangians \eqref{l1} and \eqref{l11} are invariant under the transformations involving fields swapping and change of time sign and, therefore, are $\mathcal{PT}$-invariant. However, a Lagrangian is not a physical observable (unlike a Hamiltonian), and its unclear whether the $\mathcal{PT}$-symmetry of the Lagrangian is related to real energy spectrum. To study the energy spectra, we write the Hamiltonian \eqref{hamilt2} in terms of fields and momenta as
\begin{equation}
    H=c^2\frac{\partial \phi_1}{\partial x}\,\frac{\partial \phi_2}{\partial x}+\pi_1\pi_2+\frac{1}{2\tau}\left(\pi_2\phi_2-\pi_1\phi_1\right)-\frac{1}{4\tau^2}\phi_1\phi_2
\label{Hmom}    
\end{equation}
The $\mathcal{PT}$ transformation involves changing the sign of momenta and swapping two fields. We observe that this gives $\mathcal{PT} H=H$, implying that H in (\ref{Hmom}) is $\mathcal{PT}$-symmetric and $[\mathcal{PT},H]=0$. However, this is not a sufficient condition to ensure real eigenvalues. The caveat is that the time reversal $\mathcal{T}$ is an antilinear operator. Given the eigenvectors of the Hamiltonian and $\mathcal{PT}$ operators:
\begin{equation}    H\,|n_H\rangle\,=\,\epsilon_n \,|n_H\rangle\,,\quad \quad  \mathcal{PT}\,|n_{\mathcal{PT}}\rangle\,=\,\varepsilon_n\, |n_{\mathcal{PT}}\rangle
\end{equation}
\noindent the two eigenvectors do not necessarily coincide:
\begin{equation}
   |n_H\rangle\,\neq\, |n_{\mathcal{PT}}\rangle \label{crit}
\end{equation}
This last criterion defines two different phases of the system (see Fig.\ref{fig3}): the $\mathcal{PT}$ unbroken phase and the $\mathcal{PT}$ broken phase. The first is distinguished from the second by the fact that the two sets of eigenvectors are equivalent. This implies:
\begin{align}
    &\text{$\mathcal{PT}$ unbroken phase}\,:\quad \text{real eigenvalues},\quad \mathrm{Im}(\omega_n)\,=\,0 \nonumber\\
    &\text{$\mathcal{PT}$ broken phase}\,:\quad \text{pairs of complex eigenvalues} \nonumber
\end{align}

The second phase can be viewed as the phase where coupling between the source and the sink cannot be balanced \cite{benderbook}, corresponding to a proper open system \cite{rotter2015review,davies1976quantum}. The separation between the two phases is called the ``exceptional point'' and is characterized by interesting properties such as a the halving of the degrees of freedom and unparticle physics \cite{Heiss2015,PhysRevD.86.034505}. In the context of non-Hermitian theories and dissipation, the exceptional point was discussed in Ref. \cite{10.1093/ptep/ptu183}.

The phase diagram of our system can be discussed using the eigenfrequencies of the system \eqref{compl}. They have the same form as \eqref{cc}:
\begin{equation}
    \omega\,=\,\omega_{\mathrm{Re}}\,-\,i\,\omega_{\mathrm{Im}}\label{aa}
\end{equation}
The solutions of the form \eqref{aa} are called ``quasinormal modes'' and are discussed in several fields, including dissipative open systems, holography \cite{Kovtun:2005ev}, hydrodynamics \cite{Kovtun:2012rj} and gravitational waves dynamics \cite{Nollert:1999ji,Chirenti:2017mwe}. In these areas, it is well recognised that the finite imaginary part of these modes determines the relaxation times of the excitations and governs the late-time dynamics of the physical system.

For a finite $\tau$, our system is in the $\mathcal{PT}$-broken phase because, according to Eq. (\ref{compl}), the spectrum always contains an imaginary term. The eigenvalues in our theory, given by Eq. (\ref{compl}), become real in the absence of dissipation and $\tau\rightarrow\infty$. This implies that $\tau\rightarrow\infty$ corresponds to the exceptional point. Fig. (\ref{fig3}) illustrates this point. 

As discussed in section \ref{sectwofields}, the number of degrees of freedom is halved in the absence of dissipation when $\tau\rightarrow\infty$. This is reminiscent of the exceptional point which separates the $\mathcal{PT}$ broken and unbroken phases. In our case, the exceptional point is at infinity and the system is always in the $\mathcal{PT}$ broken phase, in which the eigenvalues are complex. 

It would be interesting to deform our Lagrangian \eqref{l1} or (\ref{l11}) by adding a new scale that allows the system to cross over between two regimes in the phase diagram shown in Fig. \ref{fig3} as in simpler field theories of \cite{Alexandre:2017erl,Alexandre:2018uol,Bender:2005hf}.

\section{Correlation functions} \label{seccorr}

Using Lagrangian \eqref{l1}, the equations of motion can be written in matrix form as:
\begin{equation}
    \mathcal{K}^{ab}\,\phi_a\,=\,0
\end{equation}
where $\mathcal{K}^{ab}$ is the kinetic (matrix) operator, which in Fourier space is:
\begin{gather}
\mathcal{K}^{ab}(\omega,k)=\left(
  \begin{array}{cc}
 0&  \displaystyle   \omega^2-c^2{\bf k}^2+i\tau^{-1}\omega  \\
    \displaystyle  \omega^2-c^2{\bf k}^2-i\tau^{-1}\omega & 0\\
  \end{array}
\right)
\end{gather}
We define the matrix of Green's functions as
\begin{equation}
    G_{ab}(\omega,k)\,=\,\left[\mathcal{K}^{ab}\right]^{-1}(\omega,k)
\end{equation}
with the inverse:
\begin{gather}
G^{ab}(\omega,k)=\left(
  \begin{array}{cc}
 0&  \displaystyle   \frac{1}{ \omega^2-c^2{\bf k}^2-i\tau^{-1}\omega}  \\
    \displaystyle  \frac{1}{ \omega^2-c^2{\bf k}^2+i\tau^{-1}\omega} & 0\\
  \end{array}
\right)
\label{gmatr}
\end{gather}
Then, the correlation functions read
\begin{eqnarray}
\begin{split}
    & \langle \phi_1 \phi_2\rangle \,=\,\frac{1}{ \omega^2-c^2{\bf k}^2-i\tau^{-1}\omega}\,\label{check}\\
      & \langle \phi_2 \phi_1\rangle \,=\,\frac{1}{ \omega^2-c^2{\bf k}^2+i\tau^{-1}\omega}\,\\
      &  \langle \phi_1 \phi_1\rangle \,=\,\langle \phi_2 \phi_2\rangle\,=\,0
\end{split}     
\label{corphiom}      
    \end{eqnarray}
Taking the Fourier transform gives time dependence of these functions as
\begin{eqnarray}
\begin{split}
&\langle\phi_2{\phi_1}\rangle_t=\theta(t)\,\dfr{\sin\left(\omega_R t\right)}{\omega_R}\,e^{-\frac{t}{2\tau}}\\
&\langle\phi_1{\phi_2}\rangle_t=\theta(-t)\,\dfr{\sin\left(-\omega_R t\right)}{\omega_R}\,e^{\frac{t}{2\tau}}
\end{split}
\label{phicor}
\end{eqnarray}
where 
\begin{equation}
\omega_R=\sqrt{c^2{\bf k}^2-\frac{1}{4\tau^2}}
\label{omcorr}
\end{equation}

\noindent is the frequency related to GMS as in (\ref{gapsol}).

To find the correlation functions for fields $\psi$ fields, we use the inverse transformation of (\ref{trans1}):
\begin{eqnarray}
\begin{split}
\psi_1=\frac{1}{\sqrt{2}}(\phi_1+\phi_2)\\ \nonumber
\psi_2=\frac{1}{\sqrt{2}}(\phi_1-\phi_2)
\end{split}
\label{invtrans}
\end{eqnarray}
\noindent resulting in
\begin{eqnarray}
\begin{split}
& \langle \psi_1 \psi_1\rangle \,=\,\frac{1}{2}\left(\langle \phi_1 \phi_2\rangle\,+\,\langle \phi_2 \phi_1\rangle\right)\\ 
& \langle \psi_2 \psi_2\rangle \,=\,-\frac{1}{2}\left(\langle \phi_1 \phi_2\rangle\,+\,\langle \phi_2 \phi_1\rangle\right)\,=\,-\,\langle \psi_1 \psi_1\rangle\\
& \langle \psi_1 \psi_2\rangle \,=\,\frac{1}{2}\left(\langle \phi_2 \phi_1\rangle\,-\,\langle \phi_1 \phi_2\rangle\right)\\
& \langle \psi_2 \psi_1\rangle \,=\,\frac{1}{2}\left(\langle \phi_1 \phi_2\rangle\,-\,\langle \phi_2 \phi_1\rangle\right)=-\,\langle \psi_1 \psi_2\rangle
\end{split}
\label{trancorr}
\end{eqnarray}
Interestingly, the trace of the Green's function matrix vanishes as follows from $\langle\psi_1\psi_1\rangle=-\langle\psi_2\psi_2\rangle$ above and from (\ref{gmatr}): $\mathrm{Tr}\,G_{ab}(\omega,k)=0$.

Combining (\ref{corphiom}) and (\ref{trancorr}) gives
\begin{eqnarray}
\begin{split}
& \langle\psi_1\psi_1\rangle=-\frac{\omega^2-c^2 k^2}{(\omega^2-c^2k^2)^2+\tau^{-2}\omega^2}\\
&\langle\psi_1\psi_2\rangle=\frac{i\,\omega\,\tau^{-1}}{(\omega^2-c^2k^2)^2+\tau^{-2}\omega^2}
\end{split}
\label{psicor0}
\end{eqnarray}

Taking the Fourier transform gives time dependence of these functions as
\begin{eqnarray}
\begin{split}
     & \langle\psi_1{\psi_1}\rangle_t=\dfr{\sin\left(\omega_R|t|\right)}{2\,\omega_R}e^{-\frac{|t|}{2\tau}}\\
   &\langle\psi_1{\psi_2}\rangle_t=\dfr{\sin\left(\omega_Rt\right)}{2\omega_R}e^{-\frac{|t|}{2\tau}}
\end{split}
\label{psicorr}
\end{eqnarray}

\noindent which also follows from combining (\ref{phicor}) and (\ref{trancorr}). 

For $k>k_g$ when $\omega_R$ is real and positive, the correlators of both fields $\phi$ and $\psi$ show a damped oscillatory behavior with two important features. First, the oscillation frequency of the correlation functions is set by $\omega_R$ in (\ref{omcorr}), the frequency that sets gapped momentum states as discussed throughout this paper. This frequency features in the poles of calculated correlation functions (see, e.g., Eqs. \ref{phicor} and \ref{psicorr}). Second, the decay time of the correlation functions is set by the relaxation time $\tau$.

\begin{figure}
    \centering
    \includegraphics[width=0.9 \linewidth]{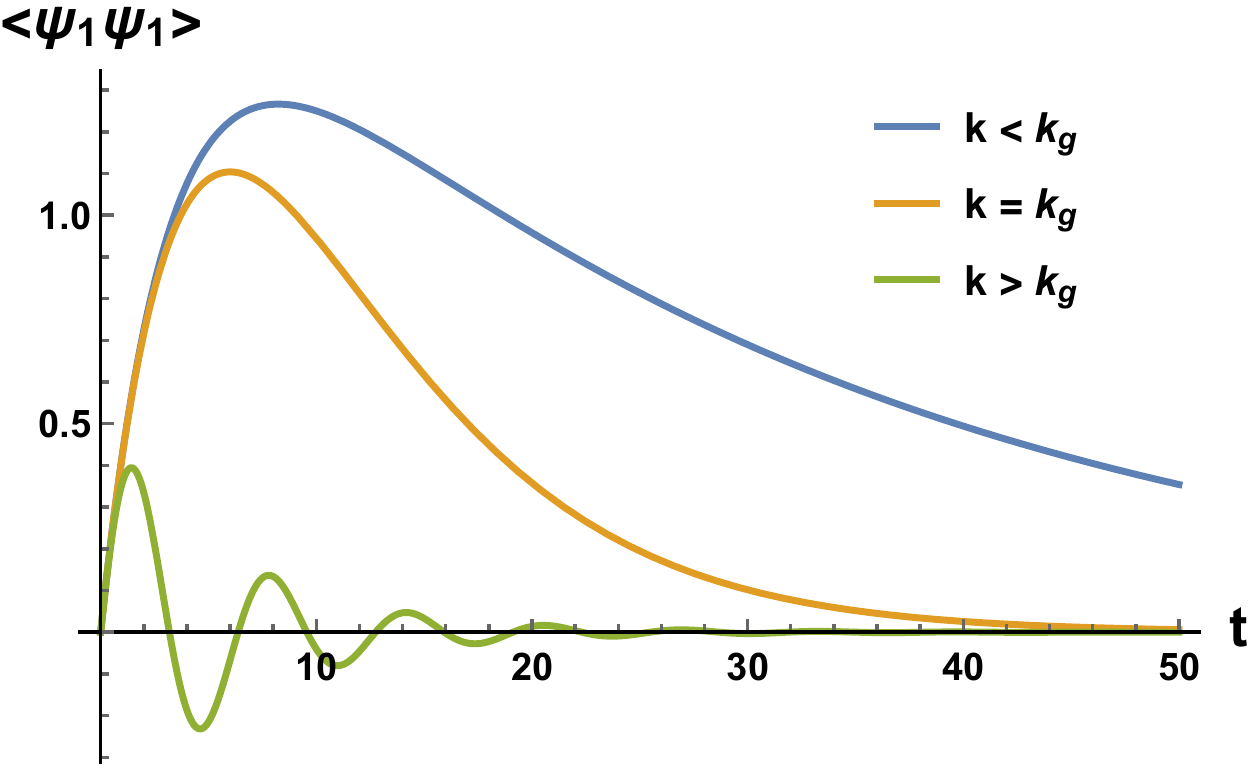}
    \caption{Plots of correlation function $\langle\psi_1{\psi_1}\rangle$ (\ref{psicorr}) in three different regimes: the solid-like regime where $k>k_g$ and where transverse modes propagate (lower curve), the hydrodynamic regime $k<k_k$ with no propagating waves (upper curve) and the exceptional point $k=k_g$ (middle curve).}
    \label{fig4}
\end{figure}

For $k<k_g$ in the fluid-mechanics regime, the oscillatory behaviour disappears, and the late time dynamics is exponentially decaying. This is illustrated in Fig. \ref{fig4} where we plot the correlation function $\langle\psi_1{\psi_1}\rangle$ (\ref{psicorr}) in three different regimes. 

\color{black}

The behavior of correlation functions in (\ref{phicor}) and (\ref{psicorr}) is expected and is physically reasonable. It shows that the non-Hermitian field theory with dissipation proposed here yields physically sensible results in terms of correlation functions and their frequency and time behavior. This is important in view of previous problems of formulating a Lagrangian-based field theory with dissipation. 

The same results for correlation functions can be obtained using path integration. The path integral has the following form:
\begin{gather}
Z=\iint \mathcal{D}\phi_1\mathcal{D}\phi_2 \,e^{-L},\quad L=\iint \mathrm{d}{\bf k}\,\mathrm{d}\omega\, \mathcal{L}_k
\end{gather}
where $\int\mathcal{D}\phi=\prod\limits_{k}\iint \mathrm{d}\phi(k)\mathrm{d}\phi(-k)$ is the functional measure, $k\equiv\{{\bf k},\,\omega\}$ the four-momentum, $\tau$ is the coherence time and $V_{\bf k}=V^{-1}$ is the system's volume in ${\bf k}$-space. If the Lagrangian has the form of \eqref{l1} then
\begin{multline}
\mathcal{L}_k=\dfr12\phi_1(k)\left(\omega^2-c^2{\bf k}^2-i\tau^{-1}\omega\right)\phi_2(-k)+\\
\dfr12\phi_2(k)\left(\omega^2-c^2{\bf k}^2+i\tau^{-1}\omega\right)\phi_1(-k)
\end{multline}
As a specific example, let us consider the calculation of the $\langle\phi_1\phi_2\rangle$ correlation function:
\begin{gather}
\langle\phi_1\phi_2\rangle_{k}=
Z^{-1}\iint \mathcal{D}\phi_1\mathcal{D}\,\phi_2 \,\phi_1(k)\phi_2(-k)\,e^{-L}
\end{gather}
Using functional derivation, the correlation function can be written as
\begin{multline}
\langle\phi_1\phi_2\rangle_{k}=\lim\limits_{\alpha\to 0}\,\dfr{\delta^2}{\delta\alpha_1(-k)\delta\alpha_1(k)}\,\times\\
\dfr{\iint \mathrm{d}\phi_1(k)\,\mathrm{d}\phi_2(-k)\,e^{-\mathcal{L}_k+\,\alpha_1(k)
\phi_1(-k)+\,\alpha_2(-k)\phi_2(k)}}{\iint \mathrm{d}{\phi_1}(k)\,\mathrm{d}\phi_2(-k)\, e^{-\mathcal{L}_k}}\,\,
\end{multline}
where $\alpha_1$ and $\alpha_2$ are the sources.
After integrating over $\phi_1$ and $\phi_2$, we obtain
\begin{align}
\langle\phi_1\phi_2\rangle_{k}&=\lim\limits_{\alpha\to 0}\,\dfr{\delta^2}{\delta\alpha_1(-k)\,\delta\alpha_1(k)}\,\exp\left(\dfr{\alpha_1(-k)\,\alpha_2(k)}{\omega^2-c^2{\bf k}^2-i\tau^{-1}\omega}\right)=\nonumber\\
&=\dfr{1}{\omega^2-c^2{\bf k}^2-i\tau^{-1}\omega}\,
\end{align}

This is the same result as $\langle\phi_1\phi_2\rangle$ in \eqref{check}. The other correlators can be obtained using the same method. This shows that the path integral formulation of our theory is sensible and gives consistent results.

\section{Interaction potential} \label{secint}
We now address the behavior of the correlation functions in space. The Fourier transform of the propagator taken in space is related to the interaction potential between particles distance $r$ apart \cite{zee}. The correlators in the presence of dissipation (\ref{corphiom})  depend on the modulus of ${\bf k}$ and are rotationally invariant. Hence, the interaction potential depends on the radial coordinate only. In order to preserve the causality of the interactions, we choose the retarded correlator in (\ref{corphiom}), whose spatial Fourier transform is
\begin{equation}
\mathfrak{D}=\frac{1}{(2\pi)^3}\int\limits_{0}^\infty\frac{e^{i\,\bf k \cdot \bf r}}{\omega^2-c^2{\bf k}^2+\frac{i\,\omega}{\tau}}\,d{\bf k}
\label{inter1}
\end{equation}
Evaluating the integral in spherical coordinates gives
\begin{equation}
\mathfrak{D}=-\frac{1}{4\pi c^2r}\,e^{\frac{i\,r}{c}\sqrt{\omega^2+\frac{i\,\omega}{\tau}}}
\label{inter2}
\end{equation}
In the absence of dissipation $\tau\rightarrow\infty$, we recover the standard result \cite{berestetskii2012quantum}
\begin{equation}
\mathfrak{D}(\tau\rightarrow\infty)=-\frac{1}{4\pi c^2r}\,e^{\frac{i\,\omega \,r}{c}}
\label{inter3}
\end{equation}
Eq. (\ref{inter2}) can be written as
\begin{equation}
\mathfrak{D}=-\frac{1}{4\pi\, c^2\,r}\,e^{\frac{i\,r}{c}\omega_1} \,e^{-\frac{r}{c}\,\omega_2}
\label{inter4}
\end{equation}

\noindent where
\begin{eqnarray}
\begin{split}
\omega_1=\frac{\omega}{\sqrt{2}}\sqrt{\sqrt{1+\left(\omega \tau\right)^{-2}}+1}\\
\omega_2=\frac{\omega}{\sqrt{2}}\sqrt{\sqrt{1+\left(\omega \tau\right)^{-2}}-1}
\end{split} 
\label{inter5}
\end{eqnarray}

In the solid-like elastic propagating regime $\omega\tau\gg 1$, $\omega_1=\omega$, $\omega_2=0$, and $\mathfrak{D}$ in Eq. (\ref{inter4}) becomes 
\begin{equation}
\mathfrak{D}=-\frac{1}{4\pi\, c^2\,r}\,e^{\frac{i\,\omega\, r}{c}}
\label{inter6}
\end{equation}

\noindent which is the same as Eq. (\ref{inter3}) in the absence of dissipation. 

In the non-propagating hydrodynamic regime $\omega\tau\ll 1$, $\omega_1=\omega_2=\sqrt{\frac{\omega}{2\tau}}$, and $\mathfrak{D}$ reads
\begin{equation}
\mathfrak{D}=-\frac{1}{4\pi\, c^2\,r}\,e^{\frac{i\,r}{c}\sqrt{\frac{\omega}{2\,\tau}}}\,e^{-\frac{r}{c}\sqrt{\frac{\omega}{2\tau}}}
\label{inter7}
\end{equation}

$\mathfrak{D}$ in Eq. (\ref{inter7}) differs from $\mathfrak{D}$ in (\ref{inter6}) in two respects. First, the oscillating part in (\ref{inter7}) can be viewed as the wave propagating with an effective frequency dependent speed $c_1$
\begin{equation}
c_1(\omega)=c\sqrt{2\,\omega\,\tau}
\end{equation}

\noindent where $c_1\ll c$ in the regime $\omega\tau\ll 1$. 

Second, $\mathfrak{D}$ in (\ref{inter7}), and therefore the corresponding interaction, become short-ranged and acquire an exponentially decreasing term $\propto\exp\left(-\frac{r}{d}\right)$, where the decay distance $d$ is
\begin{equation}
d=c\sqrt{\frac{2\,\tau}{\omega}} \label{ww}
\end{equation}

We observe that the wavelength in the oscillatory term in (\ref{inter7}) is equal to the decay distance (\ref{ww}). Therefore, $\mathfrak{D}$ displays an overdamped dependence in $r$-space in this regime, in contrast to Eq. (\ref{inter6}).

The regime $\omega \tau \ll 1$ approximately coincides with $k<k_g$ (see Eq. (\ref{omcorr})) and implies no propagating waves that mediate the interaction. It is therefore interesting that an interaction still operates and extends to a finite distance $d$ (\ref{ww}). This can be interpreted as follows. In the hydrodynamic regime $\omega \tau \ll 1$, the hydrodynamic diffusive mode is 
\begin{equation}
    \omega\,=\,-\,i\,D\,k^2\,\qquad \text{with}\qquad D\,=\,c^2\,\tau\,.
\label{ww1}    
\end{equation}

\noindent where $D$ is diffusion constant. 

Eq. (\ref{ww}) gives
\begin{equation}
    d^2\,=\,\frac{2\,c^2\,\tau}{\omega}\,=\,\frac{2\,D}{\omega}
\end{equation}

\noindent where we used (\ref{ww1}).

Using $\omega\propto\frac{1}{T}$, where $T$ is period, we find
\begin{equation}
    d^2\propto D\,T
\end{equation}

\noindent which has the form of the Einstein diffusion equation.

Physically, this implies that despite the absence of propagating waves in the regime $\omega\tau\ll 1$, the interaction is mediated by the diffusive mode up to a distance corresponding to the mean displacement in the Einstein relation. To the best of our knowledge, an interaction mediated by diffusive modes was not previously considered using formal field theory. 
 
Eq. \eqref{ww} implies that the decay distance becomes infinite in the limit of zero frequency. This corresponds to the mean displacement of the diffusive mode becoming infinite and interaction transmitted without exponential decay and screening. 

We note that $d$ in \eqref{ww} can remain finite in the limit of small $\omega$ provided $\tau$ tends to 0 (corresponding to large dissipation) in a way that the limit of the ratio $\frac{\tau}{\omega}$ in \eqref{ww} is finite.
\color{black}
\section{Further discussion}

\subsection{Departure from the ``harmonic paradigm''} \label{secharm}

Introducing the quantum field theory and its Lagrangian $L$, Zee writes \cite{zee}:
\begin{equation}
L=\frac{1}{2}\left(\sum\limits_a\,m\, \dot{q}_a^2-\sum\limits_{a,b}k_{a,b}\,q_a\,q_b-\sum\limits_{a,b,c}g_{abc}\,q_a\,q_b\,q_c-\dots \right)
\label{lzee}
\end{equation}
The first two terms in (\ref{lzee}) describe a harmonic theory and propagating plane waves, giving the starting point of the theory. The nonlinear terms describe scattering of plane waves originating from the harmonic part of the Lagrangian and production of new particles. The nonlinear terms need to be small compared to the harmonic term in order for the perturbation theory to converge and produce sensible results.

Zee observes \cite{zee} that the subject of the field theory remains rooted in this ``harmonic paradigm''. Characterising this state of affairs as limited, he wonders about ways beyond the paradigm. Notably, our approach and in particular the Lagrangian (\ref{l11}) (or (\ref{l1})) represents a departure from the harmonic paradigm in two important respects.

First, the elastic or harmonic (Klein-Gordon) term in (\ref{l11}) is not necessarily a starting point of the system description, with the viscous dissipative term added on top as is the case in the harmonic paradigm of the field theory based on (\ref{lzee}). Indeed, both elastic and viscous $\propto\frac{1}{\tau}$ term are treated in (\ref{l11}) on equal footing. The same applies to elastic and viscous terms in the Maxwell-Frenkel interpolation (\ref{a1}) where they are similarly treated on equal footing, and on which our Lagrangian formulation is based.

The combined effect of elastic and viscous terms is interestingly related to the widely-used term ``viscoelasticity'' and the area known as generalized hydrodynamics \cite{boon1991molecular,ropp}. The central effort in this area is to start with hydrodynamic equations such as Navier-Stokes equation and subsequently modify it to include the elastic response. The term and approach are related to the everyday observation that liquids flow and therefore necessitate a hydrodynamic approach as a starting point, with the elastic properties accounted for as a next step. However, we recently showed \cite{pre} that the same Eq. (\ref{gener3}) that follows from this Lagrangian can also be obtained by starting with a solid-like equation for a non-decaying wave and by subsequently generalizing the shear modulus to include the viscous response using Maxwell interpolation (\ref{a1}). Therefore, ``elastoviscosity'' is an equally legitimate term to describe Eq. (\ref{a1}) and (\ref{gener3}) as well as Lagrangians (\ref{l1}) or (\ref{l11}). This is apparent in our Lagrangian (\ref{l11}) which gives no preference as to the starting point and treats elastic and viscous terms on equal footing. 

Second and differently from (\ref{lzee}) where nonlinear terms need to be small in order for the perturbation theory to converge, the dissipative $\propto\frac{1}{\tau}$ term in (\ref{l11}) is not small in general. As discussed in the next section, large dissipative term (small $\tau$) results in purely hydrodynamic viscous regime where no shear waves propagate, completely negating the effect of the harmonic elastic term. In this sense, our approach and Lagrangian (\ref{l11}) is essentially {\it non-perturbative}. 

The crossover to non-propagating regime is similar to the concept of \textit{diffusons} which appears in the theory of electron-electron interactions in dirty metals \cite{rammer1998quantum} as well as in physics of glasses and amorphous materials \cite{doi:10.1080/13642819908223054,PhysRevLett.122.145501,PhysRevResearch.1.012010}.

Notably, we observe that all the nonlinear terms in (\ref{lzee}) can be thought to be incorporated in the dissipative $\propto\frac{1}{\tau}$ term in Lagrangian (\ref{l11}). At a deeper level, this is tantamount to stating that the introduction of liquid relaxation time $\tau$ by Frenkel \cite{frenkel} accounted for the (exponentially) complex problem of treating strongly-coupled nonlinear oscillators \cite{ropp,myprd}. \\

\subsection{Implications for a Lagrangian formulation of hydrodynamics} \label{sechydro}

As discussed earlier, the treatment of dissipative systems using a formal field theory has been a long-standing problem. A related open problem is formulating hydrodynamics, the area with a long history, using the field-theoretical description based on a Lagrangian. We note here that hydrodynamics in a broad sense is an effective field theory description valid at large wavelengths and long times. In this sense it is applicable to all systems, including crystals \cite{PhysRevA.6.2401}. In a different context, ``hydrodynamics'' is used as an equivalent to ``fluid-mechanics''and applies to liquids only \cite{landau1}. To avoid confusion, we will refer to hydrodynamics in a broad sense and to fluid dynamics in the second, more restrictive, sense.

One general approach to this problem was to use Keldysh-Schwinger-based techniques discussed earlier in this paper and out-of-time-order contours \cite{Grozdanov:2013dba,Glorioso:2018wxw,crossley,Jensen:2018hse,Haehl:2014zda} and more recently holography \cite{deBoer:2018qqm,Jana:2020vyx}. Despite the action being typically non-Hermitian (i.e. containing a finite but positive imaginary part, $\mathrm{Im}(S_{eff})>0$), a unitary evolution is ensured by the  KMS constraint \cite{Glorioso:2018wxw}. In the regime where dissipation is slow, $\tau/T \gg 1$, an interesting phenomenological formalism (not derived from an action formalism) known as \textit{quasi-hydrodynamics} \cite{Grozdanov:2018fic} has been proposed and verified explicitly in several holographic constructions \cite{Baggioli:2019sio,Baggioli:2019aqf,Baggioli:2018nnp,Baggioli:2018vfc}.

Our description of dissipation and gapped momentum states involves two fields in the Lagrangian (\ref{l1}) or (\ref{l11}) and in this sense is similar to the Keldysh-Schwinger approach where the two fields are introduced to close the integration contour (see section \ref{six}). For convenience, we re-write (\ref{l1}) and (\ref{twoeq}):
\begin{equation}
L_\phi=\frac{\partial\phi_1}{\partial t}\frac{\partial\phi_2}{\partial t}-c^2\frac{\partial\phi_1}{\partial x}\frac{\partial\phi_2}{\partial x}+\frac{1}{2\tau}\left(\phi_1\frac{\partial\phi_2}{\partial t}-\phi_2\frac{\partial\phi_1}{\partial t}\right)
\label{hydr1}
\end{equation}
\begin{eqnarray}
\begin{split}
c^2\frac{\partial^2\phi_1}{\partial x^2}=\frac{\partial^2\phi_1}{\partial t^2}+\frac{1}{\tau}\frac{\partial\phi_1}{\partial t}\\
c^2\frac{\partial^2\phi_2}{\partial x^2}=\frac{\partial^2\phi_2}{\partial t^2}-\frac{1}{\tau}\frac{\partial\phi_2}{\partial t}
\end{split}
\label{hydr2}
\end{eqnarray}
        
The Lagrangian and equations of motions have two parameters, $c$ and $\tau$. This suggests that the Lagrangian can give rise to different regimes depending on $c$ and $\tau$. Below we show that the constructed Lagrangian indeed has three well-defined regimes: (a) purely elastic non-dissipative and non-fluid regime, (b) mixed regime where transverse modes propagate above the $k$-gap and (c) purely fluid-dynamics regime where no transverse modes propagate.

As discussed earlier, the most general solution describes propagating transverse modes above the $k$-gap according to Eqs. (\ref{twosol}) and (\ref{gapsol}). This is the mixed regime (b) above. The purely elastic Lagrangian, and regime (a) above, readily follows from setting $\tau\rightarrow\infty$, in which case $\phi_1=\phi_2$, according to (\ref{twosol}), and we are left with a standard propagating Klein-Gordon field. A non-propagating regime follows from considering the condition at which fluid mechanics applies: $\omega\tau\ll 1$ \cite{frenkel,landau1}. Considering time dependence of fields $\phi_{1,2}\propto\exp(i\omega t)$, we see that the terms with second and first time derivative in (\ref{hydr2}) are $\propto\omega^2\tau^2$ and $\propto\omega\tau$, respectively. Therefore, the second time derivative term can be neglected in the regime $\omega\tau\ll 1$, and we find the ``loss'' subsystem describing the Navier-Stokes equation predicting non-propagating waves, $c^2\frac{\partial^2\phi_1}{\partial x^2}=\frac{1}{\tau}\frac{\partial\phi_1}{\partial t}$, and its gain counter-part, $c^2\frac{\partial^2\phi_2}{\partial x^2}=-\frac{1}{\tau}\frac{\partial\phi_2}{\partial t}$. 

\begin{figure}
    \centering
    \includegraphics[width=0.9 \linewidth]{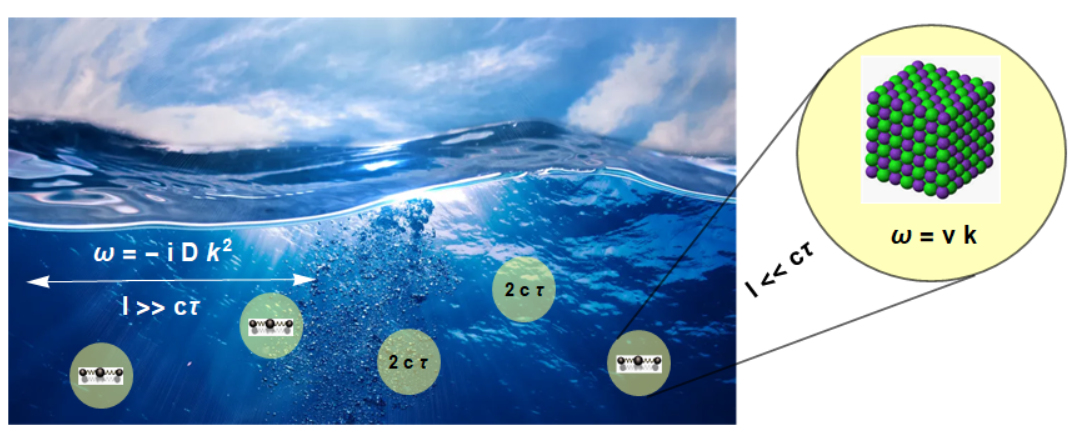}
    \caption{Illustration of different regimes described by our field theory. The field theory correctly describes a fluid system with diffusive modes only at length scales $d>2c\tau$. At smaller distances $d<2c\tau$, the field theory predicts a solid-like non-hydrodynamic behaviour with propagating shear modes.}
    \label{fig6}
\end{figure}

The different regimes also follow without bringing the frequency of external probe, $\omega$, into consideration. The product of two parameters $c$ and $\tau$ in the Lagrangian (\ref{hydr1}) gives the length scale $c\tau$. We now recall that in the fluid mechanics regime, now transverse modes operate \cite{landau1}. It is easy to see our Lagrangian describes this regime at distance 
\begin{equation}
d>2c\tau
\label{cond}
\end{equation}
Indeed, (\ref{hydr1}) results in no propagating modes when the frequencies (\ref{compl}) do not have a real part. This corresponds to $k<k_g$, or $\lambda>2c\tau$ for wavelengths (see Eq. (\ref{gapsol})). We also recall that $c\tau$ is the wave propagation range, and that a wave, in order to be well-defined, must not have a wavelength longer than the propagation range. Hence, (\ref{cond}) follows, as illustrated in Fig. \ref{fig6}.

We note that hydrodynamics and fluid mechanics are often stated to describe the medium at small $k$. The novelty here is that the field theory proposed gives a specific range of $k$ based on the parameters of the theory ($c$ and $\tau$) where the fluid-mechanical description operates. %The fact that $k_g$ indicates the breakdown of hydrodynamics is a well appreciated fact \cite{Grozdanov:2019kge,Grozdanov:2019uhi}.

We note that conditions (\ref{cond}) and $k<k_g$ are consistent with the condition of applicability of fluid mechanics discussed earlier,  $\omega\tau<1$. Indeed, combining the dispersion relation with the $k$-gap (\ref{gapsol}) with $\omega\tau<1$ gives $k<\sqrt{\frac{5}{2}}\frac{1}{c\tau}\approx\frac{1}{c\tau}$, or approximately $k<k_g$.

Increasing $\tau$ decreases the range of length scales where the hydrodynamic regime operates. At $\tau\rightarrow\infty$, this range shrinks to zero, consistent with removing the dissipative term in our lagrangian. On the other hand, small $\tau$ increases the hydrodynamic range. In this process, there is an interesting difference between the scale-free field theories and the field theory describing condensed matter phases with the UV cutoff. Unlike in scale-free field theories, $\tau$ can not decrease without bound in condensed matter phases and is bound by the shortest, Debye, vibration period, $\tau_{\rm D}$. When $\tau$ approaches $\tau_{\rm D}$, $c\tau$ becomes $c\tau_{\rm D}$ and is approximately equal to the shortest distance in condensed matter phases set by the interatomic separation $a$ (see (\ref{uv})).

It is interesting to recall that $\tau=\tau_{\rm D}$ at the UV cutoff corresponds to the Frenkel line separating the combined oscillatory and diffusive components of liquid dynamics from purely diffusive motion \cite{ropp,fr1,fr2}. $\tau=\tau_{\rm D}$ corresponds to $k_g$ approaching the zone boundary, at which point all transverse waves disappear from the system spectrum. This, in turn, corresponds to purely hydrodynamic Navier-Stokes solutions as discussed above.

We note that below the $k$-gap, Eq. \eqref{compl} gives two solutions. One of them is the diffusive mode representing the diffusive hydrodynamic shear flow:
\begin{equation}
\omega\,=\,-\,i\,c^2\,\tau\,k^2
\end{equation}
The second mode is
\begin{equation}
    \omega\,=\,-\,i\,\tau^{-1}\,+\,i\,c^2\,\tau\,k^2
\end{equation}
and is not present in the standard formulation of fluid-mechanics description \cite{landau1}.

\begin{figure}
    \centering
    \includegraphics[width=0.8\linewidth]{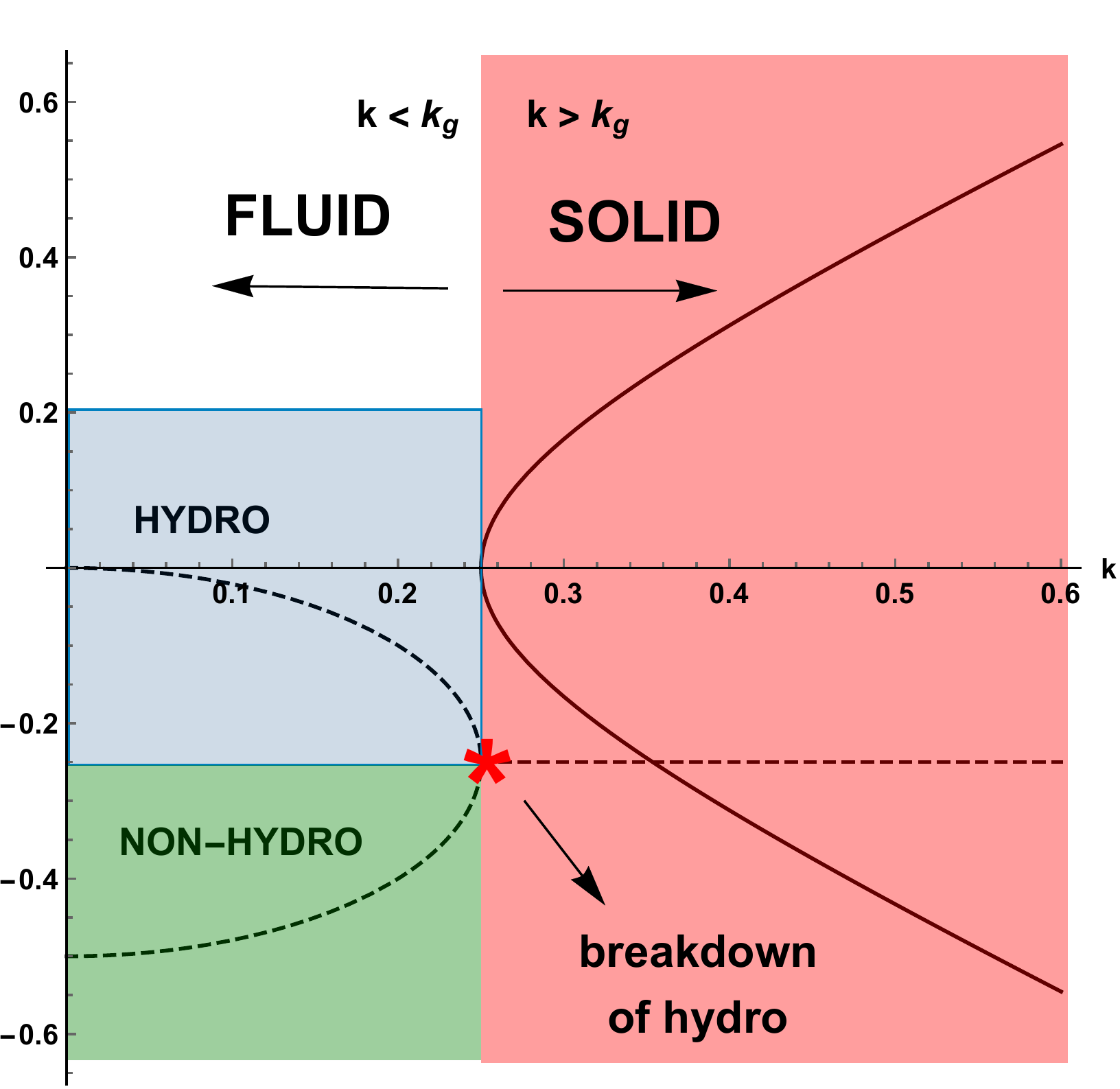}
    \caption{Different regimes of our system in relation to hydrodynamic description. The red region corresponds to solid-like propagating modes. The region below $k_g$ does not have propagating modes and we label it as the ``fluid region''. The blue area is the hydrodynamic regime $\omega \tau \ll 1\,,k c \tau \ll 1$ which is smaller than the fluid region. The red star indicates the breakdown of the hydrodynamic perturbative framework.}
    \label{ll}
\end{figure}

The second gapped mode is related to GMS, which emerges due to the collision between the diffusive and gapped modes. In a more general framework, this gapped mode can be captured by improved setups such as generalized hydrodynamics \cite{Grozdanov:2018fic}. The diffusive mode operates in the limit $\omega \tau \ll 1$ as in standard fluid mechanics, as is illustrated in Fig. \ref{ll}.

\subsection{Implications for liquid theory}

In addition to the general importance of formulating a field theory with dissipation, our results have more specific and practical interest in areas where decaying excitations are either experimentally measured or are obtained for modelling and need to fitted and analyzed. One example is the area of liquids where the measured inelastic structure factor needs to be fitted in order to extract phonon frequencies. Traditionally, the results of generalized hydrodynamics \cite{boon1991molecular,ropp} are used to fit experimental spectra. Generalized hydrodynamics starts with the hydrodynamic equations and subsequently modifies them to account for solid-like non-hydrodynamic effects such as propagating transverse waves. Traditionally, this was supported by our experience that liquids flow and hence require a hydrodynamic theory as a starting point. However, more recent experiments show that even simple low-viscous liquids such as Na, Ga, Cu, Fe and so on are not hydrodynamic but, similarly to solids, support transverse waves with frequencies approaching the highest (Debye) frequency and wavelengths approaching the interatomic separation \cite{ropp}. A hydrodynamic Navier-Stokes equation does not predict these waves \cite{landau1,frenkel}. Therefore, describing real liquids necessitates the presence of an elastic component in the equations of motion such as Eq. (\ref{gener3}) which, in turn, follows from the viscoelastic Maxwell-Frenkel theory. As discussed earlier, this elastic component is manifestly present in the Lagrangian (\ref{l11}) in the form of the Klein-Gordon fields and enters the Lagrangian on equal footing with the hydrodynamic dissipative term. This discussion therefore revisits the point discussed in the earlier section related to the correct starting point of liquid description involving both hydrodynamic and elastic terms.

Returning to the generalized hydrodynamics approach, there are issues related to its phenomenological nature and, consequently, approximations used to fit liquid spectra \cite{pilgrim,giordano}. These can affect the reliability and interpretation of experimental data. On the other hand, a Lagrangian formulation of combined effects of elasticity and dissipation is free from these complications. Moreover, the full range of field-theoretical methods can be applied to the Lagrangian to calculate different correlation functions of relevance in the liquid theory as well as other theories where dissipation plays a central role. In this respect, it is interesting to note that the functional form of the denominator of (\ref{psicor0}) is similar to that obtained in generalized hydrodynamics for correlation functions \cite{boon1991molecular}. 

\section{Conclusions}

In summary, we developed a field theory of dissipation based on gapped momentum states and using the non-Hermitian two-field theory with broken $\mathcal{PT}$ symmetry. The calculated correlation functions show decaying oscillatory behavior with the frequency and dissipation related to gapped momentum states. The interaction potential becomes short-ranged due to dissipation. We observed that the proposed field theory represents a departure from the harmonic paradigm theory and discussed the implications of this theory for the Lagrangian formulation of hydrodynamics. 

Our theory is relatively simple as compared to more complicated setups \cite{Glorioso:2018wxw} and is therefore suitable for practical and tractable calculations, providing an optimal formulation to study dissipation using field theory more generally and beyond the well-known phenomenological approaches. It would be interesting to extend our theory by adding new relevant fields and types of interaction and apply the theory to a wider range of systems of interest including, for example, electromagnetic and electron waves.

\section*{Acknowledgments}
We are grateful to J. Alexandre, D. Arean, C. Bender, M. Chernodub, A. Cortijo, N. P. Fokeeva, S. Grozdanov, K. Landsteiner, N. Poovuttikul, K. Schalm and J. Zaanen for fruitful discussions and comments. K. T. thanks EPSRC for support. M. B. acknowledges the support of the Spanish MINECO’s ``Centro de Excelencia Severo Ochoa'' Programme under grant SEV-2012-0249.

\let\oldaddcontentsline\addcontentsline% Store \addcontentsline
\renewcommand{\addcontentsline}[3]{}

\bibliographystyle{apsrev4-1}
\bibliography{field}

\begin{thebibliography}{81}%
\makeatletter
\providecommand \@ifxundefined [1]{%
 \@ifx{#1\undefined}
}%
\providecommand \@ifnum [1]{%
 \ifnum #1\expandafter \@firstoftwo
 \else \expandafter \@secondoftwo
 \fi
}%
\providecommand \@ifx [1]{%
 \ifx #1\expandafter \@firstoftwo
 \else \expandafter \@secondoftwo
 \fi
}%
\providecommand \natexlab [1]{#1}%
\providecommand \enquote  [1]{``#1''}%
\providecommand \bibnamefont  [1]{#1}%
\providecommand \bibfnamefont [1]{#1}%
\providecommand \citenamefont [1]{#1}%
\providecommand \href@noop [0]{\@secondoftwo}%
\providecommand \href [0]{\begingroup \@sanitize@url \@href}%
\providecommand \@href[1]{\@@startlink{#1}\@@href}%
\providecommand \@@href[1]{\endgroup#1\@@endlink}%
\providecommand \@sanitize@url [0]{\catcode `\\12\catcode `\$12\catcode
  `\&12\catcode `\#12\catcode `\^12\catcode `\_12\catcode `\%12\relax}%
\providecommand \@@startlink[1]{}%
\providecommand \@@endlink[0]{}%
\providecommand \url  [0]{\begingroup\@sanitize@url \@url }%
\providecommand \@url [1]{\endgroup\@href {#1}{\urlprefix }}%
\providecommand \urlprefix  [0]{URL }%
\providecommand \Eprint [0]{\href }%
\providecommand \doibase [0]{http://dx.doi.org/}%
\providecommand \selectlanguage [0]{\@gobble}%
\providecommand \bibinfo  [0]{\@secondoftwo}%
\providecommand \bibfield  [0]{\@secondoftwo}%
\providecommand \translation [1]{[#1]}%
\providecommand \BibitemOpen [0]{}%
\providecommand \bibitemStop [0]{}%
\providecommand \bibitemNoStop [0]{.\EOS\space}%
\providecommand \EOS [0]{\spacefactor3000\relax}%
\providecommand \BibitemShut  [1]{\csname bibitem#1\endcsname}%
\let\auto@bib@innerbib\@empty
%</preamble>
\bibitem [{\citenamefont {Landau}\ and\ \citenamefont
  {Lifshitz}(2013{\natexlab{a}})}]{landau}%
  \BibitemOpen
  \bibfield  {author} {\bibinfo {author} {\bibfnamefont {L.}~\bibnamefont
  {Landau}}\ and\ \bibinfo {author} {\bibfnamefont {E.}~\bibnamefont
  {Lifshitz}},\ }\href {https://books.google.es/books?id=VzgJN-XPTRsC} {\emph
  {\bibinfo {title} {Statistical Physics}}},\ \bibinfo {number} {v. 5}\
  (\bibinfo  {publisher} {Elsevier Science},\ \bibinfo {year}
  {2013})\BibitemShut {NoStop}%
\bibitem [{\citenamefont {Rotter}\ and\ \citenamefont
  {Bird}(2015)}]{rotter2015review}%
  \BibitemOpen
  \bibfield  {author} {\bibinfo {author} {\bibfnamefont {I.}~\bibnamefont
  {Rotter}}\ and\ \bibinfo {author} {\bibfnamefont {J.}~\bibnamefont {Bird}},\
  }\href@noop {} {\bibfield  {journal} {\bibinfo  {journal} {Reports on
  Progress in Physics}\ }\textbf {\bibinfo {volume} {78}},\ \bibinfo {pages}
  {114001} (\bibinfo {year} {2015})}\BibitemShut {NoStop}%
\bibitem [{\citenamefont {Bender}(2018)}]{benderbook}%
  \BibitemOpen
  \bibfield  {author} {\bibinfo {author} {\bibfnamefont {C.}~\bibnamefont
  {Bender}},\ }\href {https://books.google.es/books?id=5PF9DwAAQBAJ} {\emph
  {\bibinfo {title} {PT Symmetry: In Quantum and Classical Physics}}}\
  (\bibinfo  {publisher} {World Scientific Publishing},\ \bibinfo {year}
  {2018})\BibitemShut {NoStop}%
\bibitem [{\citenamefont {Mohsen}(2017)}]{mohsen}%
  \BibitemOpen
  \bibfield  {author} {\bibinfo {author} {\bibfnamefont {R.}~\bibnamefont
  {Mohsen}},\ }\href {https://books.google.com.fj/books?id=YfCtDgAAQBAJ} {\emph
  {\bibinfo {title} {Classical And Quantum Dissipative Systems}}}\ (\bibinfo
  {publisher} {World Scientific Publishing Company},\ \bibinfo {year}
  {2017})\BibitemShut {NoStop}%
\bibitem [{\citenamefont {Bender}(2007)}]{bender}%
  \BibitemOpen
  \bibfield  {author} {\bibinfo {author} {\bibfnamefont {C.~M.}\ \bibnamefont
  {Bender}},\ }\href {\doibase 10.1088/0034-4885/70/6/r03} {\bibfield
  {journal} {\bibinfo  {journal} {Reports on Progress in Physics}\ }\textbf
  {\bibinfo {volume} {70}},\ \bibinfo {pages} {947} (\bibinfo {year}
  {2007})}\BibitemShut {NoStop}%
\bibitem [{\citenamefont {Rotter}(2009)}]{rotter1}%
  \BibitemOpen
  \bibfield  {author} {\bibinfo {author} {\bibfnamefont {I.}~\bibnamefont
  {Rotter}},\ }\href {\doibase 10.1088/1751-8113/42/15/153001} {\bibfield
  {journal} {\bibinfo  {journal} {Journal of Physics A: Mathematical and
  Theoretical}\ }\textbf {\bibinfo {volume} {42}},\ \bibinfo {pages} {153001}
  (\bibinfo {year} {2009})}\BibitemShut {NoStop}%
\bibitem [{\citenamefont {Liu}\ and\ \citenamefont
  {Glorioso}(2018)}]{Glorioso:2018wxw}%
  \BibitemOpen
  \bibfield  {author} {\bibinfo {author} {\bibfnamefont {H.}~\bibnamefont
  {Liu}}\ and\ \bibinfo {author} {\bibfnamefont {P.}~\bibnamefont {Glorioso}},\
  }\href {\doibase 10.22323/1.305.0008} {\bibfield  {journal} {\bibinfo
  {journal} {PoS}\ }\textbf {\bibinfo {volume} {TASI2017}},\ \bibinfo {pages}
  {008} (\bibinfo {year} {2018})},\ \Eprint {http://arxiv.org/abs/1805.09331}
  {arXiv:1805.09331 [hep-th]} \BibitemShut {NoStop}%
%%CITATION = ARXIV:1805.09331;%%
\bibitem [{\citenamefont {Kamenev}(2011{\natexlab{a}})}]{kamenev}%
  \BibitemOpen
  \bibfield  {author} {\bibinfo {author} {\bibfnamefont {A.}~\bibnamefont
  {Kamenev}},\ }\href {https://books.google.es/books?id=CwlrUepnla4C} {\emph
  {\bibinfo {title} {Field Theory of Non-Equilibrium Systems}}}\ (\bibinfo
  {publisher} {Cambridge University Press},\ \bibinfo {year}
  {2011})\BibitemShut {NoStop}%
\bibitem [{\citenamefont {Endlich}\ \emph {et~al.}(2013)\citenamefont
  {Endlich}, \citenamefont {Nicolis}, \citenamefont {Porto},\ and\
  \citenamefont {Wang}}]{endlich}%
  \BibitemOpen
  \bibfield  {author} {\bibinfo {author} {\bibfnamefont {S.}~\bibnamefont
  {Endlich}}, \bibinfo {author} {\bibfnamefont {A.}~\bibnamefont {Nicolis}},
  \bibinfo {author} {\bibfnamefont {R.~A.}\ \bibnamefont {Porto}}, \ and\
  \bibinfo {author} {\bibfnamefont {J.}~\bibnamefont {Wang}},\ }\href {\doibase
  10.1103/PhysRevD.88.105001} {\bibfield  {journal} {\bibinfo  {journal} {Phys.
  Rev.}\ }\textbf {\bibinfo {volume} {D88}},\ \bibinfo {pages} {105001}
  (\bibinfo {year} {2013})},\ \Eprint {http://arxiv.org/abs/1211.6461}
  {arXiv:1211.6461 [hep-th]} \BibitemShut {NoStop}%
%%CITATION = ARXIV:1211.6461;%%
\bibitem [{\citenamefont {Crossley}\ \emph {et~al.}(2017)\citenamefont
  {Crossley}, \citenamefont {Glorioso},\ and\ \citenamefont {Liu}}]{crossley}%
  \BibitemOpen
  \bibfield  {author} {\bibinfo {author} {\bibfnamefont {M.}~\bibnamefont
  {Crossley}}, \bibinfo {author} {\bibfnamefont {P.}~\bibnamefont {Glorioso}},
  \ and\ \bibinfo {author} {\bibfnamefont {H.}~\bibnamefont {Liu}},\ }\href
  {\doibase 10.1007/JHEP09(2017)095} {\bibfield  {journal} {\bibinfo  {journal}
  {JHEP}\ }\textbf {\bibinfo {volume} {09}},\ \bibinfo {pages} {095} (\bibinfo
  {year} {2017})},\ \Eprint {http://arxiv.org/abs/1511.03646} {arXiv:1511.03646
  [hep-th]} \BibitemShut {NoStop}%
%%CITATION = ARXIV:1511.03646;%%
\bibitem [{\citenamefont {Feynman}\ and\ \citenamefont
  {Vernon}(1963)}]{feynman}%
  \BibitemOpen
  \bibfield  {author} {\bibinfo {author} {\bibfnamefont {R.}~\bibnamefont
  {Feynman}}\ and\ \bibinfo {author} {\bibfnamefont {F.}~\bibnamefont
  {Vernon}},\ }\href {\doibase https://doi.org/10.1016/0003-4916(63)90068-X}
  {\bibfield  {journal} {\bibinfo  {journal} {Annals of Physics}\ }\textbf
  {\bibinfo {volume} {24}},\ \bibinfo {pages} {118 } (\bibinfo {year}
  {1963})}\BibitemShut {NoStop}%
\bibitem [{\citenamefont {Caldeira}\ and\ \citenamefont
  {Leggett}(1981)}]{leggett}%
  \BibitemOpen
  \bibfield  {author} {\bibinfo {author} {\bibfnamefont {A.~O.}\ \bibnamefont
  {Caldeira}}\ and\ \bibinfo {author} {\bibfnamefont {A.~J.}\ \bibnamefont
  {Leggett}},\ }\href {\doibase 10.1103/PhysRevLett.46.211} {\bibfield
  {journal} {\bibinfo  {journal} {Phys. Rev. Lett.}\ }\textbf {\bibinfo
  {volume} {46}},\ \bibinfo {pages} {211} (\bibinfo {year} {1981})}\BibitemShut
  {NoStop}%
\bibitem [{\citenamefont {Weiss}(1999)}]{weiss}%
  \BibitemOpen
  \bibfield  {author} {\bibinfo {author} {\bibfnamefont {U.}~\bibnamefont
  {Weiss}},\ }\href {https://books.google.es/books?id=kqZclKUZdq0C} {\emph
  {\bibinfo {title} {Quantum Dissipative Systems}}},\ Series in modern
  condensed matter physics\ (\bibinfo  {publisher} {World Scientific},\
  \bibinfo {year} {1999})\BibitemShut {NoStop}%
\bibitem [{\citenamefont {Baggioli}(2019)}]{Baggioli:2019rrs}%
  \BibitemOpen
  \bibfield  {author} {\bibinfo {author} {\bibfnamefont {M.}~\bibnamefont
  {Baggioli}},\ }\emph {\bibinfo {title} {{Applied Holography}}},\ \href
  {\doibase 10.1007/978-3-030-35184-7} {Ph.D. thesis},\ \bibinfo  {school}
  {Madrid, IFT} (\bibinfo {year} {2019}),\ \Eprint
  {http://arxiv.org/abs/1908.02667} {arXiv:1908.02667 [hep-th]} \BibitemShut
  {NoStop}%
%%CITATION = ARXIV:1908.02667;%%
\bibitem [{\citenamefont {Policastro}\ \emph {et~al.}(2002)\citenamefont
  {Policastro}, \citenamefont {Son},\ and\ \citenamefont
  {Starinets}}]{Policastro:2002se}%
  \BibitemOpen
  \bibfield  {author} {\bibinfo {author} {\bibfnamefont {G.}~\bibnamefont
  {Policastro}}, \bibinfo {author} {\bibfnamefont {D.~T.}\ \bibnamefont {Son}},
  \ and\ \bibinfo {author} {\bibfnamefont {A.~O.}\ \bibnamefont {Starinets}},\
  }\href {\doibase 10.1088/1126-6708/2002/09/043} {\bibfield  {journal}
  {\bibinfo  {journal} {JHEP}\ }\textbf {\bibinfo {volume} {09}},\ \bibinfo
  {pages} {043} (\bibinfo {year} {2002})},\ \Eprint
  {http://arxiv.org/abs/hep-th/0205052} {arXiv:hep-th/0205052 [hep-th]}
  \BibitemShut {NoStop}%
%%CITATION = HEP-TH/0205052;%%
\bibitem [{\citenamefont {Janik}(2007)}]{Janik:2006ft}%
  \BibitemOpen
  \bibfield  {author} {\bibinfo {author} {\bibfnamefont {R.~A.}\ \bibnamefont
  {Janik}},\ }\href {\doibase 10.1103/PhysRevLett.98.022302} {\bibfield
  {journal} {\bibinfo  {journal} {Phys. Rev. Lett.}\ }\textbf {\bibinfo
  {volume} {98}},\ \bibinfo {pages} {022302} (\bibinfo {year} {2007})},\
  \Eprint {http://arxiv.org/abs/hep-th/0610144} {arXiv:hep-th/0610144 [hep-th]}
  \BibitemShut {NoStop}%
%%CITATION = HEP-TH/0610144;%%
\bibitem [{\citenamefont {Baggioli}\ and\ \citenamefont
  {Trachenko}(2019{\natexlab{a}})}]{Baggioli:2018vfc}%
  \BibitemOpen
  \bibfield  {author} {\bibinfo {author} {\bibfnamefont {M.}~\bibnamefont
  {Baggioli}}\ and\ \bibinfo {author} {\bibfnamefont {K.}~\bibnamefont
  {Trachenko}},\ }\href {\doibase 10.1007/JHEP03(2019)093} {\bibfield
  {journal} {\bibinfo  {journal} {JHEP}\ }\textbf {\bibinfo {volume} {03}},\
  \bibinfo {pages} {093} (\bibinfo {year} {2019}{\natexlab{a}})},\ \Eprint
  {http://arxiv.org/abs/1807.10530} {arXiv:1807.10530 [hep-th]} \BibitemShut
  {NoStop}%
%%CITATION = ARXIV:1807.10530;%%
\bibitem [{\citenamefont {Baggioli}\ and\ \citenamefont
  {Trachenko}(2019{\natexlab{b}})}]{Baggioli:2018nnp}%
  \BibitemOpen
  \bibfield  {author} {\bibinfo {author} {\bibfnamefont {M.}~\bibnamefont
  {Baggioli}}\ and\ \bibinfo {author} {\bibfnamefont {K.}~\bibnamefont
  {Trachenko}},\ }\href {\doibase 10.1103/PhysRevD.99.106002} {\bibfield
  {journal} {\bibinfo  {journal} {Phys. Rev.}\ }\textbf {\bibinfo {volume}
  {D99}},\ \bibinfo {pages} {106002} (\bibinfo {year} {2019}{\natexlab{b}})},\
  \Eprint {http://arxiv.org/abs/1808.05391} {arXiv:1808.05391 [hep-th]}
  \BibitemShut {NoStop}%
%%CITATION = ARXIV:1808.05391;%%
\bibitem [{\citenamefont {Baggioli}\ \emph
  {et~al.}(2019{\natexlab{a}})\citenamefont {Baggioli}, \citenamefont {Gran},
  \citenamefont {Alba}, \citenamefont {Tornsö},\ and\ \citenamefont
  {Zingg}}]{Baggioli:2019aqf}%
  \BibitemOpen
  \bibfield  {author} {\bibinfo {author} {\bibfnamefont {M.}~\bibnamefont
  {Baggioli}}, \bibinfo {author} {\bibfnamefont {U.}~\bibnamefont {Gran}},
  \bibinfo {author} {\bibfnamefont {A.~J.}\ \bibnamefont {Alba}}, \bibinfo
  {author} {\bibfnamefont {M.}~\bibnamefont {Tornsö}}, \ and\ \bibinfo
  {author} {\bibfnamefont {T.}~\bibnamefont {Zingg}},\ }\href {\doibase
  10.1007/JHEP09(2019)013} {\bibfield  {journal} {\bibinfo  {journal} {JHEP}\
  }\textbf {\bibinfo {volume} {09}},\ \bibinfo {pages} {013} (\bibinfo {year}
  {2019}{\natexlab{a}})},\ \Eprint {http://arxiv.org/abs/1905.00804}
  {arXiv:1905.00804 [hep-th]} \BibitemShut {NoStop}%
%%CITATION = ARXIV:1905.00804;%%
\bibitem [{\citenamefont {Baggioli}\ \emph
  {et~al.}(2019{\natexlab{b}})\citenamefont {Baggioli}, \citenamefont {Gran},\
  and\ \citenamefont {Tornsö}}]{Baggioli:2019sio}%
  \BibitemOpen
  \bibfield  {author} {\bibinfo {author} {\bibfnamefont {M.}~\bibnamefont
  {Baggioli}}, \bibinfo {author} {\bibfnamefont {U.}~\bibnamefont {Gran}}, \
  and\ \bibinfo {author} {\bibfnamefont {M.}~\bibnamefont {Tornsö}},\
  }\href@noop {} {\  (\bibinfo {year} {2019}{\natexlab{b}})},\ \Eprint
  {http://arxiv.org/abs/1912.07321} {arXiv:1912.07321 [hep-th]} \BibitemShut
  {NoStop}%
%%CITATION = ARXIV:1912.07321;%%
\bibitem [{\citenamefont {Trachenko}\ and\ \citenamefont
  {Brazhkin}(2015)}]{ropp}%
  \BibitemOpen
  \bibfield  {author} {\bibinfo {author} {\bibfnamefont {K.}~\bibnamefont
  {Trachenko}}\ and\ \bibinfo {author} {\bibfnamefont {V.~V.}\ \bibnamefont
  {Brazhkin}},\ }\href {\doibase 10.1088/0034-4885/79/1/016502} {\bibfield
  {journal} {\bibinfo  {journal} {Reports on Progress in Physics}\ }\textbf
  {\bibinfo {volume} {79}},\ \bibinfo {pages} {016502} (\bibinfo {year}
  {2015})}\BibitemShut {NoStop}%
\bibitem [{\citenamefont {Yang}\ \emph {et~al.}(2017)\citenamefont {Yang},
  \citenamefont {Dove}, \citenamefont {Brazhkin},\ and\ \citenamefont
  {Trachenko}}]{prl}%
  \BibitemOpen
  \bibfield  {author} {\bibinfo {author} {\bibfnamefont {C.}~\bibnamefont
  {Yang}}, \bibinfo {author} {\bibfnamefont {M.~T.}\ \bibnamefont {Dove}},
  \bibinfo {author} {\bibfnamefont {V.~V.}\ \bibnamefont {Brazhkin}}, \ and\
  \bibinfo {author} {\bibfnamefont {K.}~\bibnamefont {Trachenko}},\ }\href
  {\doibase 10.1103/PhysRevLett.118.215502} {\bibfield  {journal} {\bibinfo
  {journal} {Phys. Rev. Lett.}\ }\textbf {\bibinfo {volume} {118}},\ \bibinfo
  {pages} {215502} (\bibinfo {year} {2017})}\BibitemShut {NoStop}%
\bibitem [{\citenamefont {Trachenko}(2017{\natexlab{a}})}]{pre}%
  \BibitemOpen
  \bibfield  {author} {\bibinfo {author} {\bibfnamefont {K.}~\bibnamefont
  {Trachenko}},\ }\href {\doibase 10.1103/PhysRevE.96.062134} {\bibfield
  {journal} {\bibinfo  {journal} {Phys. Rev. E}\ }\textbf {\bibinfo {volume}
  {96}},\ \bibinfo {pages} {062134} (\bibinfo {year}
  {2017}{\natexlab{a}})}\BibitemShut {NoStop}%
\bibitem [{\citenamefont {Trachenko}\ and\ \citenamefont
  {Brazhkin}(2014)}]{collective}%
  \BibitemOpen
  \bibfield  {author} {\bibinfo {author} {\bibfnamefont {K.}~\bibnamefont
  {Trachenko}}\ and\ \bibinfo {author} {\bibfnamefont {V.}~\bibnamefont
  {Brazhkin}},\ }\href@noop {} {\bibfield  {journal} {\bibinfo  {journal}
  {Journal of Physical Chemistry B}\ }\textbf {\bibinfo {volume} {118}},\
  \bibinfo {pages} {11417} (\bibinfo {year} {2014})}\BibitemShut {NoStop}%
\bibitem [{\citenamefont {Baggioli}\ and\ \citenamefont
  {Zaccone}(2019{\natexlab{a}})}]{PhysRevLett.122.145501}%
  \BibitemOpen
  \bibfield  {author} {\bibinfo {author} {\bibfnamefont {M.}~\bibnamefont
  {Baggioli}}\ and\ \bibinfo {author} {\bibfnamefont {A.}~\bibnamefont
  {Zaccone}},\ }\href {\doibase 10.1103/PhysRevLett.122.145501} {\bibfield
  {journal} {\bibinfo  {journal} {Phys. Rev. Lett.}\ }\textbf {\bibinfo
  {volume} {122}},\ \bibinfo {pages} {145501} (\bibinfo {year}
  {2019}{\natexlab{a}})}\BibitemShut {NoStop}%
\bibitem [{\citenamefont {Baggioli}\ \emph
  {et~al.}(2019{\natexlab{c}})\citenamefont {Baggioli}, \citenamefont
  {Milkus},\ and\ \citenamefont {Zaccone}}]{PhysRevE.100.062131}%
  \BibitemOpen
  \bibfield  {author} {\bibinfo {author} {\bibfnamefont {M.}~\bibnamefont
  {Baggioli}}, \bibinfo {author} {\bibfnamefont {R.}~\bibnamefont {Milkus}}, \
  and\ \bibinfo {author} {\bibfnamefont {A.}~\bibnamefont {Zaccone}},\ }\href
  {\doibase 10.1103/PhysRevE.100.062131} {\bibfield  {journal} {\bibinfo
  {journal} {Phys. Rev. E}\ }\textbf {\bibinfo {volume} {100}},\ \bibinfo
  {pages} {062131} (\bibinfo {year} {2019}{\natexlab{c}})}\BibitemShut
  {NoStop}%
\bibitem [{\citenamefont {Baggioli}\ and\ \citenamefont
  {Zaccone}(2019{\natexlab{b}})}]{PhysRevResearch.1.012010}%
  \BibitemOpen
  \bibfield  {author} {\bibinfo {author} {\bibfnamefont {M.}~\bibnamefont
  {Baggioli}}\ and\ \bibinfo {author} {\bibfnamefont {A.}~\bibnamefont
  {Zaccone}},\ }\href {\doibase 10.1103/PhysRevResearch.1.012010} {\bibfield
  {journal} {\bibinfo  {journal} {Phys. Rev. Research}\ }\textbf {\bibinfo
  {volume} {1}},\ \bibinfo {pages} {012010} (\bibinfo {year}
  {2019}{\natexlab{b}})}\BibitemShut {NoStop}%
\bibitem [{\citenamefont {Baggioli}\ and\ \citenamefont
  {Zaccone}(2020)}]{PhysRevResearch.2.013267}%
  \BibitemOpen
  \bibfield  {author} {\bibinfo {author} {\bibfnamefont {M.}~\bibnamefont
  {Baggioli}}\ and\ \bibinfo {author} {\bibfnamefont {A.}~\bibnamefont
  {Zaccone}},\ }\href {\doibase 10.1103/PhysRevResearch.2.013267} {\bibfield
  {journal} {\bibinfo  {journal} {Phys. Rev. Research}\ }\textbf {\bibinfo
  {volume} {2}},\ \bibinfo {pages} {013267} (\bibinfo {year}
  {2020})}\BibitemShut {NoStop}%
\bibitem [{\citenamefont {Baggioli}\ \emph {et~al.}(2020)\citenamefont
  {Baggioli}, \citenamefont {Vasin}, \citenamefont {Brazhkin},\ and\
  \citenamefont {Trachenko}}]{Baggioli:2019jcm}%
  \BibitemOpen
  \bibfield  {author} {\bibinfo {author} {\bibfnamefont {M.}~\bibnamefont
  {Baggioli}}, \bibinfo {author} {\bibfnamefont {M.}~\bibnamefont {Vasin}},
  \bibinfo {author} {\bibfnamefont {V.}~\bibnamefont {Brazhkin}}, \ and\
  \bibinfo {author} {\bibfnamefont {K.}~\bibnamefont {Trachenko}},\ }\href
  {\doibase https://doi.org/10.1016/j.physrep.2020.04.002} {\bibfield
  {journal} {\bibinfo  {journal} {Physics Reports}\ } (\bibinfo {year}
  {2020}),\ https://doi.org/10.1016/j.physrep.2020.04.002}\BibitemShut
  {NoStop}%
\bibitem [{\citenamefont {Zee}(2003)}]{zee}%
  \BibitemOpen
  \bibfield  {author} {\bibinfo {author} {\bibfnamefont {A.}~\bibnamefont
  {Zee}},\ }\href {https://books.google.es/books?id=85G9QgAACAAJ} {\emph
  {\bibinfo {title} {Quantum Field Theory in a Nutshell}}}\ (\bibinfo
  {publisher} {Princeton University Press},\ \bibinfo {year}
  {2003})\BibitemShut {NoStop}%
\bibitem [{\citenamefont {Maxwell}(1867)}]{maxwell}%
  \BibitemOpen
  \bibfield  {author} {\bibinfo {author} {\bibfnamefont {J.~C.}\ \bibnamefont
  {Maxwell}},\ }\href {\doibase 10.1098/rstl.1867.0004} {\bibfield  {journal}
  {\bibinfo  {journal} {Philosophical Transactions of the Royal Society of
  London}\ }\textbf {\bibinfo {volume} {157}},\ \bibinfo {pages} {49} (\bibinfo
  {year} {1867})},\ \Eprint
  {http://arxiv.org/abs/https://royalsocietypublishing.org/doi/pdf/10.1098/rstl.1867.0004}
  {https://royalsocietypublishing.org/doi/pdf/10.1098/rstl.1867.0004}
  \BibitemShut {NoStop}%
\bibitem [{\citenamefont {Frenkel}(1955)}]{frenkel}%
  \BibitemOpen
  \bibfield  {author} {\bibinfo {author} {\bibfnamefont {J.}~\bibnamefont
  {Frenkel}},\ }\href {https://books.google.es/books?id=ORdSQwAACAAJ} {\emph
  {\bibinfo {title} {Kinetic Theory of Liquids}}},\ Dover Publications\
  (\bibinfo  {publisher} {Dover},\ \bibinfo {year} {1955})\BibitemShut
  {NoStop}%
\bibitem [{\citenamefont {Dyre}(2006)}]{dyre}%
  \BibitemOpen
  \bibfield  {author} {\bibinfo {author} {\bibfnamefont {J.~C.}\ \bibnamefont
  {Dyre}},\ }\href {\doibase 10.1103/RevModPhys.78.953} {\bibfield  {journal}
  {\bibinfo  {journal} {Rev. Mod. Phys.}\ }\textbf {\bibinfo {volume} {78}},\
  \bibinfo {pages} {953} (\bibinfo {year} {2006})}\BibitemShut {NoStop}%
\bibitem [{\citenamefont {Trachenko}\ and\ \citenamefont
  {Brazhkin}(2009)}]{elength}%
  \BibitemOpen
  \bibfield  {author} {\bibinfo {author} {\bibfnamefont {K.}~\bibnamefont
  {Trachenko}}\ and\ \bibinfo {author} {\bibfnamefont {V.~V.}\ \bibnamefont
  {Brazhkin}},\ }\href {\doibase 10.1088/0953-8984/21/42/425104} {\bibfield
  {journal} {\bibinfo  {journal} {Journal of Physics: Condensed Matter}\
  }\textbf {\bibinfo {volume} {21}},\ \bibinfo {pages} {425104} (\bibinfo
  {year} {2009})}\BibitemShut {NoStop}%
\bibitem [{\citenamefont {Feinberg}(1967)}]{tachyons}%
  \BibitemOpen
  \bibfield  {author} {\bibinfo {author} {\bibfnamefont {G.}~\bibnamefont
  {Feinberg}},\ }\href {\doibase 10.1103/PhysRev.159.1089} {\bibfield
  {journal} {\bibinfo  {journal} {Phys. Rev.}\ }\textbf {\bibinfo {volume}
  {159}},\ \bibinfo {pages} {1089} (\bibinfo {year} {1967})}\BibitemShut
  {NoStop}%
\bibitem [{\citenamefont {Brazhkin}\ and\ \citenamefont
  {Trachenko}(2012)}]{fr1}%
  \BibitemOpen
  \bibfield  {author} {\bibinfo {author} {\bibfnamefont {V.~V.}\ \bibnamefont
  {Brazhkin}}\ and\ \bibinfo {author} {\bibfnamefont {K.}~\bibnamefont
  {Trachenko}},\ }\href {\doibase 10.1063/PT.3.1796} {\bibfield  {journal}
  {\bibinfo  {journal} {Physics Today}\ }\textbf {\bibinfo {volume} {65}},\
  \bibinfo {pages} {68} (\bibinfo {year} {2012})},\ \Eprint
  {http://arxiv.org/abs/https://doi.org/10.1063/PT.3.1796}
  {https://doi.org/10.1063/PT.3.1796} \BibitemShut {NoStop}%
\bibitem [{\citenamefont {Brazhkin}\ \emph {et~al.}(2013)\citenamefont
  {Brazhkin}, \citenamefont {Fomin}, \citenamefont {Lyapin}, \citenamefont
  {Ryzhov}, \citenamefont {Tsiok},\ and\ \citenamefont {Trachenko}}]{fr2}%
  \BibitemOpen
  \bibfield  {author} {\bibinfo {author} {\bibfnamefont {V.~V.}\ \bibnamefont
  {Brazhkin}}, \bibinfo {author} {\bibfnamefont {Y.~D.}\ \bibnamefont {Fomin}},
  \bibinfo {author} {\bibfnamefont {A.~G.}\ \bibnamefont {Lyapin}}, \bibinfo
  {author} {\bibfnamefont {V.~N.}\ \bibnamefont {Ryzhov}}, \bibinfo {author}
  {\bibfnamefont {E.~N.}\ \bibnamefont {Tsiok}}, \ and\ \bibinfo {author}
  {\bibfnamefont {K.}~\bibnamefont {Trachenko}},\ }\href {\doibase
  10.1103/PhysRevLett.111.145901} {\bibfield  {journal} {\bibinfo  {journal}
  {Phys. Rev. Lett.}\ }\textbf {\bibinfo {volume} {111}},\ \bibinfo {pages}
  {145901} (\bibinfo {year} {2013})}\BibitemShut {NoStop}%
\bibitem [{\citenamefont {Trachenko}\ and\ \citenamefont
  {Brazhkin}(2020)}]{minimal}%
  \BibitemOpen
  \bibfield  {author} {\bibinfo {author} {\bibfnamefont {K.}~\bibnamefont
  {Trachenko}}\ and\ \bibinfo {author} {\bibfnamefont {V.}~\bibnamefont
  {Brazhkin}},\ }\href {\doibase 10.1126/sciadv.aba3747} {\bibfield  {journal}
  {\bibinfo  {journal} {Science Advances}\ }\textbf {\bibinfo {volume} {6}},\
  \bibinfo {pages} {eaba3747} (\bibinfo {year} {2020})}\BibitemShut {NoStop}%
\bibitem [{\citenamefont {Trachenko}\ \emph {et~al.}(2020)\citenamefont
  {Trachenko}, \citenamefont {Brazhkin},\ and\ \citenamefont
  {Baggioli}}]{Baggioli:2020lcf}%
  \BibitemOpen
  \bibfield  {author} {\bibinfo {author} {\bibfnamefont {K.}~\bibnamefont
  {Trachenko}}, \bibinfo {author} {\bibfnamefont {V.}~\bibnamefont {Brazhkin}},
  \ and\ \bibinfo {author} {\bibfnamefont {M.}~\bibnamefont {Baggioli}},\
  }\href@noop {} {\  (\bibinfo {year} {2020})},\ \Eprint
  {http://arxiv.org/abs/2003.13506} {arXiv:2003.13506 [hep-th]} \BibitemShut
  {NoStop}%
%%CITATION = ARXIV:2003.13506;%%
\bibitem [{\citenamefont {Trachenko}(2019)}]{myscirep}%
  \BibitemOpen
  \bibfield  {author} {\bibinfo {author} {\bibfnamefont {K.}~\bibnamefont
  {Trachenko}},\ }\href {\doibase 10.1038/s41598-019-43273-9} {\bibfield
  {journal} {\bibinfo  {journal} {Scientific Reports}\ }\textbf {\bibinfo
  {volume} {9}},\ \bibinfo {pages} {6766} (\bibinfo {year} {2019})}\BibitemShut
  {NoStop}%
\bibitem [{\citenamefont {Bateman}(1931)}]{bateman}%
  \BibitemOpen
  \bibfield  {author} {\bibinfo {author} {\bibfnamefont {H.}~\bibnamefont
  {Bateman}},\ }\href {\doibase 10.1103/PhysRev.38.815} {\bibfield  {journal}
  {\bibinfo  {journal} {Phys. Rev.}\ }\textbf {\bibinfo {volume} {38}},\
  \bibinfo {pages} {815} (\bibinfo {year} {1931})}\BibitemShut {NoStop}%
\bibitem [{\citenamefont {Dekker}(1981)}]{dekker}%
  \BibitemOpen
  \bibfield  {author} {\bibinfo {author} {\bibfnamefont {H.}~\bibnamefont
  {Dekker}},\ }\href {\doibase https://doi.org/10.1016/0370-1573(81)90033-8}
  {\bibfield  {journal} {\bibinfo  {journal} {Physics Reports}\ }\textbf
  {\bibinfo {volume} {80}},\ \bibinfo {pages} {1 } (\bibinfo {year}
  {1981})}\BibitemShut {NoStop}%
\bibitem [{\citenamefont {Landau}\ and\ \citenamefont
  {Lifshitz}(2013{\natexlab{b}})}]{landau1}%
  \BibitemOpen
  \bibfield  {author} {\bibinfo {author} {\bibfnamefont {L.}~\bibnamefont
  {Landau}}\ and\ \bibinfo {author} {\bibfnamefont {E.}~\bibnamefont
  {Lifshitz}},\ }\href {https://books.google.es/books?id=CeBbAwAAQBAJ} {\emph
  {\bibinfo {title} {Fluid Mechanics}}},\ \bibinfo {number} {v. 6}\ (\bibinfo
  {publisher} {Elsevier Science},\ \bibinfo {year} {2013})\BibitemShut
  {NoStop}%
\bibitem [{\citenamefont {Martin}\ \emph {et~al.}(1972)\citenamefont {Martin},
  \citenamefont {Parodi},\ and\ \citenamefont {Pershan}}]{PhysRevA.6.2401}%
  \BibitemOpen
  \bibfield  {author} {\bibinfo {author} {\bibfnamefont {P.~C.}\ \bibnamefont
  {Martin}}, \bibinfo {author} {\bibfnamefont {O.}~\bibnamefont {Parodi}}, \
  and\ \bibinfo {author} {\bibfnamefont {P.~S.}\ \bibnamefont {Pershan}},\
  }\href {\doibase 10.1103/PhysRevA.6.2401} {\bibfield  {journal} {\bibinfo
  {journal} {Phys. Rev. A}\ }\textbf {\bibinfo {volume} {6}},\ \bibinfo {pages}
  {2401} (\bibinfo {year} {1972})}\BibitemShut {NoStop}%
\bibitem [{\citenamefont {Ammon}\ \emph {et~al.}(2020)\citenamefont {Ammon},
  \citenamefont {Baggioli}, \citenamefont {Gray}, \citenamefont {Grieninger},\
  and\ \citenamefont {Jain}}]{Ammon:2020xyv}%
  \BibitemOpen
  \bibfield  {author} {\bibinfo {author} {\bibfnamefont {M.}~\bibnamefont
  {Ammon}}, \bibinfo {author} {\bibfnamefont {M.}~\bibnamefont {Baggioli}},
  \bibinfo {author} {\bibfnamefont {S.}~\bibnamefont {Gray}}, \bibinfo {author}
  {\bibfnamefont {S.}~\bibnamefont {Grieninger}}, \ and\ \bibinfo {author}
  {\bibfnamefont {A.}~\bibnamefont {Jain}},\ }\href@noop {} {\  (\bibinfo
  {year} {2020})},\ \Eprint {http://arxiv.org/abs/2001.05737} {arXiv:2001.05737
  [hep-th]} \BibitemShut {NoStop}%
%%CITATION = ARXIV:2001.05737;%%
\bibitem [{\citenamefont {Alexandre}\ \emph {et~al.}(2019)\citenamefont
  {Alexandre}, \citenamefont {Ellis}, \citenamefont {Millington},\ and\
  \citenamefont {Seynaeve}}]{alexandre1}%
  \BibitemOpen
  \bibfield  {author} {\bibinfo {author} {\bibfnamefont {J.}~\bibnamefont
  {Alexandre}}, \bibinfo {author} {\bibfnamefont {J.}~\bibnamefont {Ellis}},
  \bibinfo {author} {\bibfnamefont {P.}~\bibnamefont {Millington}}, \ and\
  \bibinfo {author} {\bibfnamefont {D.}~\bibnamefont {Seynaeve}},\ }\href
  {\doibase 10.1103/PhysRevD.99.075024} {\bibfield  {journal} {\bibinfo
  {journal} {Phys. Rev.}\ }\textbf {\bibinfo {volume} {D99}},\ \bibinfo {pages}
  {075024} (\bibinfo {year} {2019})},\ \Eprint
  {http://arxiv.org/abs/1808.00944} {arXiv:1808.00944 [hep-th]} \BibitemShut
  {NoStop}%
%%CITATION = ARXIV:1808.00944;%%
\bibitem [{\citenamefont {Alexandre}\ \emph {et~al.}(2017)\citenamefont
  {Alexandre}, \citenamefont {Millington},\ and\ \citenamefont
  {Seynaeve}}]{alexandre2}%
  \BibitemOpen
  \bibfield  {author} {\bibinfo {author} {\bibfnamefont {J.}~\bibnamefont
  {Alexandre}}, \bibinfo {author} {\bibfnamefont {P.}~\bibnamefont
  {Millington}}, \ and\ \bibinfo {author} {\bibfnamefont {D.}~\bibnamefont
  {Seynaeve}},\ }\href {\doibase 10.1103/PhysRevD.96.065027} {\bibfield
  {journal} {\bibinfo  {journal} {Phys. Rev.}\ }\textbf {\bibinfo {volume}
  {D96}},\ \bibinfo {pages} {065027} (\bibinfo {year} {2017})},\ \Eprint
  {http://arxiv.org/abs/1707.01057} {arXiv:1707.01057 [hep-th]} \BibitemShut
  {NoStop}%
%%CITATION = ARXIV:1707.01057;%%
\bibitem [{\citenamefont {Bender}\ \emph
  {et~al.}(2005{\natexlab{a}})\citenamefont {Bender}, \citenamefont {Brandt},
  \citenamefont {Chen},\ and\ \citenamefont {Wang}}]{Bender:2004sv}%
  \BibitemOpen
  \bibfield  {author} {\bibinfo {author} {\bibfnamefont {C.~M.}\ \bibnamefont
  {Bender}}, \bibinfo {author} {\bibfnamefont {S.~F.}\ \bibnamefont {Brandt}},
  \bibinfo {author} {\bibfnamefont {J.-H.}\ \bibnamefont {Chen}}, \ and\
  \bibinfo {author} {\bibfnamefont {Q.-h.}\ \bibnamefont {Wang}},\ }\href
  {\doibase 10.1103/PhysRevD.71.025014} {\bibfield  {journal} {\bibinfo
  {journal} {Phys. Rev.}\ }\textbf {\bibinfo {volume} {D71}},\ \bibinfo {pages}
  {025014} (\bibinfo {year} {2005}{\natexlab{a}})},\ \Eprint
  {http://arxiv.org/abs/hep-th/0411064} {arXiv:hep-th/0411064 [hep-th]}
  \BibitemShut {NoStop}%
%%CITATION = HEP-TH/0411064;%%
\bibitem [{\citenamefont {Mannheim}(2013)}]{Mannheim:2009zj}%
  \BibitemOpen
  \bibfield  {author} {\bibinfo {author} {\bibfnamefont {P.~D.}\ \bibnamefont
  {Mannheim}},\ }\href {\doibase 10.1098/rsta.2012.0060} {\bibfield  {journal}
  {\bibinfo  {journal} {Phil. Trans. Roy. Soc. Lond. A}\ }\textbf {\bibinfo
  {volume} {371}},\ \bibinfo {pages} {20120060} (\bibinfo {year} {2013})},\
  \Eprint {http://arxiv.org/abs/0912.2635} {arXiv:0912.2635 [hep-th]}
  \BibitemShut {NoStop}%
\bibitem [{\citenamefont {Kamenev}(2011{\natexlab{b}})}]{kamenev_2011}%
  \BibitemOpen
  \bibfield  {author} {\bibinfo {author} {\bibfnamefont {A.}~\bibnamefont
  {Kamenev}},\ }\href {\doibase 10.1017/CBO9781139003667} {\emph {\bibinfo
  {title} {Field Theory of Non-Equilibrium Systems}}}\ (\bibinfo  {publisher}
  {Cambridge University Press},\ \bibinfo {year} {2011})\BibitemShut {NoStop}%
\bibitem [{\citenamefont {Sudarshan}\ \emph {et~al.}(1978)\citenamefont
  {Sudarshan}, \citenamefont {Chiu},\ and\ \citenamefont {Gorini}}]{sudarshan}%
  \BibitemOpen
  \bibfield  {author} {\bibinfo {author} {\bibfnamefont {E.~C.~G.}\
  \bibnamefont {Sudarshan}}, \bibinfo {author} {\bibfnamefont {C.~B.}\
  \bibnamefont {Chiu}}, \ and\ \bibinfo {author} {\bibfnamefont
  {V.}~\bibnamefont {Gorini}},\ }\href {\doibase 10.1103/PhysRevD.18.2914}
  {\bibfield  {journal} {\bibinfo  {journal} {Phys. Rev. D}\ }\textbf {\bibinfo
  {volume} {18}},\ \bibinfo {pages} {2914} (\bibinfo {year}
  {1978})}\BibitemShut {NoStop}%
\bibitem [{\citenamefont {Sudarshan}(2010)}]{sud1}%
  \BibitemOpen
  \bibfield  {author} {\bibinfo {author} {\bibfnamefont {E.~G.}\ \bibnamefont
  {Sudarshan}},\ }\href {\doibase 10.1143/PTPS.184.451} {\bibfield  {journal}
  {\bibinfo  {journal} {Progress of Theoretical Physics Supplement}\ }\textbf
  {\bibinfo {volume} {184}},\ \bibinfo {pages} {451} (\bibinfo {year}
  {2010})},\ \Eprint
  {http://arxiv.org/abs/https://academic.oup.com/ptps/article-pdf/doi/10.1143/PTPS.184.451/5254474/184-451.pdf}
  {https://academic.oup.com/ptps/article-pdf/doi/10.1143/PTPS.184.451/5254474/184-451.pdf}
  \BibitemShut {NoStop}%
\bibitem [{\citenamefont {Feinberg}(2011)}]{Feinberg:2010xw}%
  \BibitemOpen
  \bibfield  {author} {\bibinfo {author} {\bibfnamefont {J.}~\bibnamefont
  {Feinberg}},\ }\href {\doibase 10.1007/s10773-010-0604-y} {\bibfield
  {journal} {\bibinfo  {journal} {International Journal of Theoretical
  Physics}\ }\textbf {\bibinfo {volume} {50}},\ \bibinfo {pages} {1116}
  (\bibinfo {year} {2011})}\BibitemShut {NoStop}%
\bibitem [{\citenamefont {Chernodub}\ and\ \citenamefont
  {Cortijo}(2019)}]{Chernodub:2019ggz}%
  \BibitemOpen
  \bibfield  {author} {\bibinfo {author} {\bibfnamefont {M.~N.}\ \bibnamefont
  {Chernodub}}\ and\ \bibinfo {author} {\bibfnamefont {A.}~\bibnamefont
  {Cortijo}},\ }\href@noop {} {\  (\bibinfo {year} {2019})},\ \Eprint
  {http://arxiv.org/abs/1901.06167} {arXiv:1901.06167 [cond-mat.mes-hall]}
  \BibitemShut {NoStop}%
%%CITATION = ARXIV:1901.06167;%%
\bibitem [{\citenamefont
  {Mostafazadeh}(2002{\natexlab{a}})}]{Mostafazadeh:2001jk}%
  \BibitemOpen
  \bibfield  {author} {\bibinfo {author} {\bibfnamefont {A.}~\bibnamefont
  {Mostafazadeh}},\ }\href {\doibase 10.1063/1.1418246} {\bibfield  {journal}
  {\bibinfo  {journal} {J. Math. Phys.}\ }\textbf {\bibinfo {volume} {43}},\
  \bibinfo {pages} {205} (\bibinfo {year} {2002}{\natexlab{a}})},\ \Eprint
  {http://arxiv.org/abs/math-ph/0107001} {arXiv:math-ph/0107001 [math-ph]}
  \BibitemShut {NoStop}%
%%CITATION = MATH-PH/0107001;%%
\bibitem [{\citenamefont
  {Mostafazadeh}(2002{\natexlab{b}})}]{Mostafazadeh:2001nr}%
  \BibitemOpen
  \bibfield  {author} {\bibinfo {author} {\bibfnamefont {A.}~\bibnamefont
  {Mostafazadeh}},\ }\href {\doibase 10.1063/1.1461427} {\bibfield  {journal}
  {\bibinfo  {journal} {J. Math. Phys.}\ }\textbf {\bibinfo {volume} {43}},\
  \bibinfo {pages} {2814} (\bibinfo {year} {2002}{\natexlab{b}})},\ \Eprint
  {http://arxiv.org/abs/math-ph/0110016} {arXiv:math-ph/0110016 [math-ph]}
  \BibitemShut {NoStop}%
%%CITATION = MATH-PH/0110016;%%
\bibitem [{\citenamefont
  {Mostafazadeh}(2002{\natexlab{c}})}]{Mostafazadeh:2002id}%
  \BibitemOpen
  \bibfield  {author} {\bibinfo {author} {\bibfnamefont {A.}~\bibnamefont
  {Mostafazadeh}},\ }\href {\doibase 10.1063/1.1489072} {\bibfield  {journal}
  {\bibinfo  {journal} {J. Math. Phys.}\ }\textbf {\bibinfo {volume} {43}},\
  \bibinfo {pages} {3944} (\bibinfo {year} {2002}{\natexlab{c}})},\ \Eprint
  {http://arxiv.org/abs/math-ph/0203005} {arXiv:math-ph/0203005 [math-ph]}
  \BibitemShut {NoStop}%
%%CITATION = MATH-PH/0203005;%%
\bibitem [{\citenamefont {Davies}(1976)}]{davies1976quantum}%
  \BibitemOpen
  \bibfield  {author} {\bibinfo {author} {\bibfnamefont {E.}~\bibnamefont
  {Davies}},\ }\href {https://books.google.es/books?id=nsh-AAAAIAAJ} {\emph
  {\bibinfo {title} {Quantum theory of open systems}}}\ (\bibinfo  {publisher}
  {Academic Press},\ \bibinfo {year} {1976})\BibitemShut {NoStop}%
\bibitem [{\citenamefont {Heiss}(2015)}]{Heiss2015}%
  \BibitemOpen
  \bibfield  {author} {\bibinfo {author} {\bibfnamefont {W.~D.}\ \bibnamefont
  {Heiss}},\ }\href {\doibase 10.1007/s10773-014-2428-7} {\bibfield  {journal}
  {\bibinfo  {journal} {International Journal of Theoretical Physics}\ }\textbf
  {\bibinfo {volume} {54}},\ \bibinfo {pages} {3954} (\bibinfo {year}
  {2015})}\BibitemShut {NoStop}%
\bibitem [{\citenamefont {Zubkov}(2012)}]{PhysRevD.86.034505}%
  \BibitemOpen
  \bibfield  {author} {\bibinfo {author} {\bibfnamefont {M.~A.}\ \bibnamefont
  {Zubkov}},\ }\href {\doibase 10.1103/PhysRevD.86.034505} {\bibfield
  {journal} {\bibinfo  {journal} {Phys. Rev. D}\ }\textbf {\bibinfo {volume}
  {86}},\ \bibinfo {pages} {034505} (\bibinfo {year} {2012})}\BibitemShut
  {NoStop}%
\bibitem [{\citenamefont {Hashimoto}\ \emph {et~al.}(2015)\citenamefont
  {Hashimoto}, \citenamefont {Kanki}, \citenamefont {Hayakawa},\ and\
  \citenamefont {Petrosky}}]{10.1093/ptep/ptu183}%
  \BibitemOpen
  \bibfield  {author} {\bibinfo {author} {\bibfnamefont {K.}~\bibnamefont
  {Hashimoto}}, \bibinfo {author} {\bibfnamefont {K.}~\bibnamefont {Kanki}},
  \bibinfo {author} {\bibfnamefont {H.}~\bibnamefont {Hayakawa}}, \ and\
  \bibinfo {author} {\bibfnamefont {T.}~\bibnamefont {Petrosky}},\ }\href@noop
  {} {\bibfield  {journal} {\bibinfo  {journal} {Progress of Theoretical and
  Experimental Physics}\ }\textbf {\bibinfo {volume} {2015}},\ \bibinfo {pages}
  {023A02} (\bibinfo {year} {2015})}\BibitemShut {NoStop}%
\bibitem [{\citenamefont {Kovtun}\ and\ \citenamefont
  {Starinets}(2005)}]{Kovtun:2005ev}%
  \BibitemOpen
  \bibfield  {author} {\bibinfo {author} {\bibfnamefont {P.~K.}\ \bibnamefont
  {Kovtun}}\ and\ \bibinfo {author} {\bibfnamefont {A.~O.}\ \bibnamefont
  {Starinets}},\ }\href {\doibase 10.1103/PhysRevD.72.086009} {\bibfield
  {journal} {\bibinfo  {journal} {Phys. Rev.}\ }\textbf {\bibinfo {volume}
  {D72}},\ \bibinfo {pages} {086009} (\bibinfo {year} {2005})},\ \Eprint
  {http://arxiv.org/abs/hep-th/0506184} {arXiv:hep-th/0506184 [hep-th]}
  \BibitemShut {NoStop}%
%%CITATION = HEP-TH/0506184;%%
\bibitem [{\citenamefont {Kovtun}(2012)}]{Kovtun:2012rj}%
  \BibitemOpen
  \bibfield  {author} {\bibinfo {author} {\bibfnamefont {P.}~\bibnamefont
  {Kovtun}},\ }\bibfield  {booktitle} {\emph {\bibinfo {booktitle} {{Lectures
  on hydrodynamic fluctuations in relativistic theories}}},\ }\href {\doibase
  10.1088/1751-8113/45/47/473001} {\bibfield  {journal} {\bibinfo  {journal}
  {J. Phys.}\ }\textbf {\bibinfo {volume} {A45}},\ \bibinfo {pages} {473001}
  (\bibinfo {year} {2012})},\ \Eprint {http://arxiv.org/abs/1205.5040}
  {arXiv:1205.5040 [hep-th]} \BibitemShut {NoStop}%
%%CITATION = ARXIV:1205.5040;%%
\bibitem [{\citenamefont {Nollert}(1999)}]{Nollert:1999ji}%
  \BibitemOpen
  \bibfield  {author} {\bibinfo {author} {\bibfnamefont {H.-P.}\ \bibnamefont
  {Nollert}},\ }\href {\doibase 10.1088/0264-9381/16/12/201} {\bibfield
  {journal} {\bibinfo  {journal} {Class. Quant. Grav.}\ }\textbf {\bibinfo
  {volume} {16}},\ \bibinfo {pages} {R159} (\bibinfo {year}
  {1999})}\BibitemShut {NoStop}%
%%CITATION = CQGRD,16,R159;%%
\bibitem [{\citenamefont {Chirenti}(2018)}]{Chirenti:2017mwe}%
  \BibitemOpen
  \bibfield  {author} {\bibinfo {author} {\bibfnamefont {C.}~\bibnamefont
  {Chirenti}},\ }\bibfield  {booktitle} {\emph {\bibinfo {booktitle} {{Black
  hole quasinormal modes in the era of LIGO}}},\ }\href {\doibase
  10.1007/s13538-017-0543-7} {\bibfield  {journal} {\bibinfo  {journal} {Braz.
  J. Phys.}\ }\textbf {\bibinfo {volume} {48}},\ \bibinfo {pages} {102}
  (\bibinfo {year} {2018})},\ \Eprint {http://arxiv.org/abs/1708.04476}
  {arXiv:1708.04476 [gr-qc]} \BibitemShut {NoStop}%
%%CITATION = ARXIV:1708.04476;%%
\bibitem [{\citenamefont {Alexandre}\ \emph
  {et~al.}(2018{\natexlab{a}})\citenamefont {Alexandre}, \citenamefont
  {Millington},\ and\ \citenamefont {Seynaeve}}]{Alexandre:2017erl}%
  \BibitemOpen
  \bibfield  {author} {\bibinfo {author} {\bibfnamefont {J.}~\bibnamefont
  {Alexandre}}, \bibinfo {author} {\bibfnamefont {P.}~\bibnamefont
  {Millington}}, \ and\ \bibinfo {author} {\bibfnamefont {D.}~\bibnamefont
  {Seynaeve}},\ }\bibfield  {booktitle} {\emph {\bibinfo {booktitle} {{}}},\
  }\href {\doibase 10.1088/1742-6596/952/1/012012} {\bibfield  {journal}
  {\bibinfo  {journal} {J. Phys. Conf. Ser.}\ }\textbf {\bibinfo {volume}
  {952}},\ \bibinfo {pages} {012012} (\bibinfo {year} {2018}{\natexlab{a}})},\
  \Eprint {http://arxiv.org/abs/1710.01076} {arXiv:1710.01076 [hep-th]}
  \BibitemShut {NoStop}%
%%CITATION = ARXIV:1710.01076;%%
\bibitem [{\citenamefont {Alexandre}\ \emph
  {et~al.}(2018{\natexlab{b}})\citenamefont {Alexandre}, \citenamefont {Ellis},
  \citenamefont {Millington},\ and\ \citenamefont
  {Seynaeve}}]{Alexandre:2018uol}%
  \BibitemOpen
  \bibfield  {author} {\bibinfo {author} {\bibfnamefont {J.}~\bibnamefont
  {Alexandre}}, \bibinfo {author} {\bibfnamefont {J.}~\bibnamefont {Ellis}},
  \bibinfo {author} {\bibfnamefont {P.}~\bibnamefont {Millington}}, \ and\
  \bibinfo {author} {\bibfnamefont {D.}~\bibnamefont {Seynaeve}},\ }\href
  {\doibase 10.1103/PhysRevD.98.045001} {\bibfield  {journal} {\bibinfo
  {journal} {Phys. Rev.}\ }\textbf {\bibinfo {volume} {D98}},\ \bibinfo {pages}
  {045001} (\bibinfo {year} {2018}{\natexlab{b}})},\ \Eprint
  {http://arxiv.org/abs/1805.06380} {arXiv:1805.06380 [hep-th]} \BibitemShut
  {NoStop}%
%%CITATION = ARXIV:1805.06380;%%
\bibitem [{\citenamefont {Bender}\ \emph
  {et~al.}(2005{\natexlab{b}})\citenamefont {Bender}, \citenamefont {Jones},\
  and\ \citenamefont {Rivers}}]{Bender:2005hf}%
  \BibitemOpen
  \bibfield  {author} {\bibinfo {author} {\bibfnamefont {C.~M.}\ \bibnamefont
  {Bender}}, \bibinfo {author} {\bibfnamefont {H.}~\bibnamefont {Jones}}, \
  and\ \bibinfo {author} {\bibfnamefont {R.}~\bibnamefont {Rivers}},\ }\href
  {\doibase https://doi.org/10.1016/j.physletb.2005.08.087} {\bibfield
  {journal} {\bibinfo  {journal} {Physics Letters B}\ }\textbf {\bibinfo
  {volume} {625}},\ \bibinfo {pages} {333 } (\bibinfo {year}
  {2005}{\natexlab{b}})}\BibitemShut {NoStop}%
\bibitem [{\citenamefont {Berestetskii}\ \emph {et~al.}(2012)\citenamefont
  {Berestetskii}, \citenamefont {Pitaevskii},\ and\ \citenamefont
  {Lifshitz}}]{berestetskii2012quantum}%
  \BibitemOpen
  \bibfield  {author} {\bibinfo {author} {\bibfnamefont {V.}~\bibnamefont
  {Berestetskii}}, \bibinfo {author} {\bibfnamefont {L.}~\bibnamefont
  {Pitaevskii}}, \ and\ \bibinfo {author} {\bibfnamefont {E.}~\bibnamefont
  {Lifshitz}},\ }\href {https://books.google.es/books?id=Tpk-lqyr3GoC} {\emph
  {\bibinfo {title} {Quantum Electrodynamics: Volume 4}}},\ \bibinfo {number}
  {v. 4}\ (\bibinfo  {publisher} {Elsevier Science},\ \bibinfo {year}
  {2012})\BibitemShut {NoStop}%
\bibitem [{\citenamefont {Boon}\ and\ \citenamefont
  {Yip}(1991)}]{boon1991molecular}%
  \BibitemOpen
  \bibfield  {author} {\bibinfo {author} {\bibfnamefont {J.}~\bibnamefont
  {Boon}}\ and\ \bibinfo {author} {\bibfnamefont {S.}~\bibnamefont {Yip}},\
  }\href {https://books.google.es/books?id=GcTR1xSl0ZkC} {\emph {\bibinfo
  {title} {Molecular Hydrodynamics}}},\ Dover books on physics\ (\bibinfo
  {publisher} {Dover Publications},\ \bibinfo {year} {1991})\BibitemShut
  {NoStop}%
\bibitem [{\citenamefont {Rammer}(1998)}]{rammer1998quantum}%
  \BibitemOpen
  \bibfield  {author} {\bibinfo {author} {\bibfnamefont {J.}~\bibnamefont
  {Rammer}},\ }\href {https://books.google.es/books?id=GGYsAAAAYAAJ} {\emph
  {\bibinfo {title} {Quantum Transport Theory}}},\ Frontiers in physics\
  (\bibinfo  {publisher} {Perseus Books},\ \bibinfo {year} {1998})\BibitemShut
  {NoStop}%
\bibitem [{\citenamefont {Allen}\ \emph {et~al.}(1999)\citenamefont {Allen},
  \citenamefont {Feldman}, \citenamefont {Fabian},\ and\ \citenamefont
  {Wooten}}]{doi:10.1080/13642819908223054}%
  \BibitemOpen
  \bibfield  {author} {\bibinfo {author} {\bibfnamefont {P.~B.}\ \bibnamefont
  {Allen}}, \bibinfo {author} {\bibfnamefont {J.~L.}\ \bibnamefont {Feldman}},
  \bibinfo {author} {\bibfnamefont {J.}~\bibnamefont {Fabian}}, \ and\ \bibinfo
  {author} {\bibfnamefont {F.}~\bibnamefont {Wooten}},\ }\href {\doibase
  10.1080/13642819908223054} {\bibfield  {journal} {\bibinfo  {journal}
  {Philosophical Magazine B}\ }\textbf {\bibinfo {volume} {79}},\ \bibinfo
  {pages} {1715} (\bibinfo {year} {1999})},\ \Eprint
  {http://arxiv.org/abs/https://doi.org/10.1080/13642819908223054}
  {https://doi.org/10.1080/13642819908223054} \BibitemShut {NoStop}%
\bibitem [{\citenamefont {Trachenko}(2017{\natexlab{b}})}]{myprd}%
  \BibitemOpen
  \bibfield  {author} {\bibinfo {author} {\bibfnamefont {K.}~\bibnamefont
  {Trachenko}},\ }\href {\doibase 10.1103/PhysRevD.95.043522} {\bibfield
  {journal} {\bibinfo  {journal} {Phys. Rev. D}\ }\textbf {\bibinfo {volume}
  {95}},\ \bibinfo {pages} {043522} (\bibinfo {year}
  {2017}{\natexlab{b}})}\BibitemShut {NoStop}%
\bibitem [{\citenamefont {Grozdanov}\ and\ \citenamefont
  {Polonyi}(2015)}]{Grozdanov:2013dba}%
  \BibitemOpen
  \bibfield  {author} {\bibinfo {author} {\bibfnamefont {S.}~\bibnamefont
  {Grozdanov}}\ and\ \bibinfo {author} {\bibfnamefont {J.}~\bibnamefont
  {Polonyi}},\ }\href {\doibase 10.1103/PhysRevD.91.105031} {\bibfield
  {journal} {\bibinfo  {journal} {Phys. Rev.}\ }\textbf {\bibinfo {volume}
  {D91}},\ \bibinfo {pages} {105031} (\bibinfo {year} {2015})},\ \Eprint
  {http://arxiv.org/abs/1305.3670} {arXiv:1305.3670 [hep-th]} \BibitemShut
  {NoStop}%
%%CITATION = ARXIV:1305.3670;%%
\bibitem [{\citenamefont {Jensen}\ \emph {et~al.}(2018)\citenamefont {Jensen},
  \citenamefont {Marjieh}, \citenamefont {Pinzani-Fokeeva},\ and\ \citenamefont
  {Yarom}}]{Jensen:2018hse}%
  \BibitemOpen
  \bibfield  {author} {\bibinfo {author} {\bibfnamefont {K.}~\bibnamefont
  {Jensen}}, \bibinfo {author} {\bibfnamefont {R.}~\bibnamefont {Marjieh}},
  \bibinfo {author} {\bibfnamefont {N.}~\bibnamefont {Pinzani-Fokeeva}}, \ and\
  \bibinfo {author} {\bibfnamefont {A.}~\bibnamefont {Yarom}},\ }\href
  {\doibase 10.21468/SciPostPhys.5.5.053} {\bibfield  {journal} {\bibinfo
  {journal} {SciPost Phys.}\ }\textbf {\bibinfo {volume} {5}},\ \bibinfo
  {pages} {053} (\bibinfo {year} {2018})},\ \Eprint
  {http://arxiv.org/abs/1804.04654} {arXiv:1804.04654 [hep-th]} \BibitemShut
  {NoStop}%
%%CITATION = ARXIV:1804.04654;%%
\bibitem [{\citenamefont {Haehl}\ \emph {et~al.}(2015)\citenamefont {Haehl},
  \citenamefont {Loganayagam},\ and\ \citenamefont
  {Rangamani}}]{Haehl:2014zda}%
  \BibitemOpen
  \bibfield  {author} {\bibinfo {author} {\bibfnamefont {F.~M.}\ \bibnamefont
  {Haehl}}, \bibinfo {author} {\bibfnamefont {R.}~\bibnamefont {Loganayagam}},
  \ and\ \bibinfo {author} {\bibfnamefont {M.}~\bibnamefont {Rangamani}},\
  }\href {\doibase 10.1103/PhysRevLett.114.201601} {\bibfield  {journal}
  {\bibinfo  {journal} {Phys. Rev. Lett.}\ }\textbf {\bibinfo {volume} {114}},\
  \bibinfo {pages} {201601} (\bibinfo {year} {2015})},\ \Eprint
  {http://arxiv.org/abs/1412.1090} {arXiv:1412.1090 [hep-th]} \BibitemShut
  {NoStop}%
%%CITATION = ARXIV:1412.1090;%%
\bibitem [{\citenamefont {de~Boer}\ \emph {et~al.}(2019)\citenamefont
  {de~Boer}, \citenamefont {Heller},\ and\ \citenamefont
  {Pinzani-Fokeeva}}]{deBoer:2018qqm}%
  \BibitemOpen
  \bibfield  {author} {\bibinfo {author} {\bibfnamefont {J.}~\bibnamefont
  {de~Boer}}, \bibinfo {author} {\bibfnamefont {M.~P.}\ \bibnamefont {Heller}},
  \ and\ \bibinfo {author} {\bibfnamefont {N.}~\bibnamefont
  {Pinzani-Fokeeva}},\ }\href {\doibase 10.1007/JHEP05(2019)188} {\bibfield
  {journal} {\bibinfo  {journal} {JHEP}\ }\textbf {\bibinfo {volume} {05}},\
  \bibinfo {pages} {188} (\bibinfo {year} {2019})},\ \Eprint
  {http://arxiv.org/abs/1812.06093} {arXiv:1812.06093 [hep-th]} \BibitemShut
  {NoStop}%
%%CITATION = ARXIV:1812.06093;%%
\bibitem [{\citenamefont {Jana}\ \emph {et~al.}(2020)\citenamefont {Jana},
  \citenamefont {Loganayagam},\ and\ \citenamefont {Rangamani}}]{Jana:2020vyx}%
  \BibitemOpen
  \bibfield  {author} {\bibinfo {author} {\bibfnamefont {C.}~\bibnamefont
  {Jana}}, \bibinfo {author} {\bibfnamefont {R.}~\bibnamefont {Loganayagam}}, \
  and\ \bibinfo {author} {\bibfnamefont {M.}~\bibnamefont {Rangamani}},\
  }\href@noop {} {\  (\bibinfo {year} {2020})},\ \Eprint
  {http://arxiv.org/abs/2004.02888} {arXiv:2004.02888 [hep-th]} \BibitemShut
  {NoStop}%
%%CITATION = ARXIV:2004.02888;%%
\bibitem [{\citenamefont {Grozdanov}\ \emph {et~al.}(2019)\citenamefont
  {Grozdanov}, \citenamefont {Lucas},\ and\ \citenamefont
  {Poovuttikul}}]{Grozdanov:2018fic}%
  \BibitemOpen
  \bibfield  {author} {\bibinfo {author} {\bibfnamefont {S.}~\bibnamefont
  {Grozdanov}}, \bibinfo {author} {\bibfnamefont {A.}~\bibnamefont {Lucas}}, \
  and\ \bibinfo {author} {\bibfnamefont {N.}~\bibnamefont {Poovuttikul}},\
  }\href {\doibase 10.1103/PhysRevD.99.086012} {\bibfield  {journal} {\bibinfo
  {journal} {Phys. Rev.}\ }\textbf {\bibinfo {volume} {D99}},\ \bibinfo {pages}
  {086012} (\bibinfo {year} {2019})},\ \Eprint
  {http://arxiv.org/abs/1810.10016} {arXiv:1810.10016 [hep-th]} \BibitemShut
  {NoStop}%
%%CITATION = ARXIV:1810.10016;%%
\bibitem [{\citenamefont {Pilgrim}\ and\ \citenamefont
  {Morkel}(2006)}]{pilgrim}%
  \BibitemOpen
  \bibfield  {author} {\bibinfo {author} {\bibfnamefont {W.-C.}\ \bibnamefont
  {Pilgrim}}\ and\ \bibinfo {author} {\bibfnamefont {C.}~\bibnamefont
  {Morkel}},\ }\href {\doibase 10.1088/0953-8984/18/37/r01} {\bibfield
  {journal} {\bibinfo  {journal} {Journal of Physics: Condensed Matter}\
  }\textbf {\bibinfo {volume} {18}},\ \bibinfo {pages} {R585} (\bibinfo {year}
  {2006})}\BibitemShut {NoStop}%
\bibitem [{\citenamefont {Giordano}\ and\ \citenamefont
  {Monaco}(2010)}]{giordano}%
  \BibitemOpen
  \bibfield  {author} {\bibinfo {author} {\bibfnamefont {V.~M.}\ \bibnamefont
  {Giordano}}\ and\ \bibinfo {author} {\bibfnamefont {G.}~\bibnamefont
  {Monaco}},\ }\href {\doibase 10.1073/pnas.1006319107} {\bibfield  {journal}
  {\bibinfo  {journal} {Proceedings of the National Academy of Sciences}\
  }\textbf {\bibinfo {volume} {107}},\ \bibinfo {pages} {21985} (\bibinfo
  {year} {2010})},\ \Eprint
  {http://arxiv.org/abs/https://www.pnas.org/content/107/51/21985.full.pdf}
  {https://www.pnas.org/content/107/51/21985.full.pdf} \BibitemShut {NoStop}%
\end{thebibliography}%

\appendix
\end{document}